%% file: bare_conf.tex
\documentclass[conference]{IEEEtran}
 \makeatletter
 \def\ps@headings{%
 \def\@oddhead{\mbox{}\scriptsize\rightmark \hfil \thepage}%
 \def\@evenhead{\scriptsize\thepage \hfil \leftmark\mbox{}}%
 \def\@oddfoot{}%
 \def\@evenfoot{}}
 \makeatother
\ifCLASSINFOpdf
\else
\fi
\usepackage{epsfig}
\usepackage{subfigure}
\usepackage{amsmath}
\usepackage{bm}
\usepackage{graphicx}
\usepackage{cite}
\usepackage{enumerate}
\usepackage{algorithm}
\usepackage{algorithmic}
\usepackage{balance}
\usepackage{graphicx}
\usepackage{footmisc}
\usepackage{graphicx}
\usepackage{amsmath}
\usepackage{url}
\usepackage{amsbsy}
\usepackage{bm}
\usepackage{amssymb,amsmath}
\usepackage{amsfonts}
\usepackage{array,cases}
\usepackage{paralist}
\usepackage[top = 0.75in,
bottom = 1.043in,
left = 0.639in,
right = 0.639in]{geometry}
\input{mymath_mod_JJS.tex}

\newcommand{\argmin}{\operatornamewithlimits{arg\ min}}

\begin{document}
\title{Delay-Optimal Data Forwarding in Vehicular Sensor Networks}

\author{
\IEEEauthorblockN{Okyoung Choi\IEEEauthorrefmark{2}, Seokhyun Kim\IEEEauthorrefmark{2}, Jaeseong Jeong\IEEEauthorrefmark{2}, Hyang-Won Lee\IEEEauthorrefmark{3} and Song Chong\IEEEauthorrefmark{2}}
\IEEEauthorblockA{\IEEEauthorrefmark{2}Department of Electrical Engineering, KAIST
\\ \IEEEauthorrefmark{3}Department of Internet and Multimedia Engineering, Konkuk University
% \\E-mail: \{okyoung,kimseokhyun,jsjeong\}@netsys.kaist.ac.kr, songchong@kaist.edu
}
}


%Telephone: (800) 555--1212, Fax: (888) 555--1212}


\maketitle

%\begin{abstract}
%\boldmath
%The abstract goes here.
%\end{abstract}


% \fi
\IEEEpeerreviewmaketitle

%\section{Introduction}
%This demo file is intended to serve as a ``starter file''
%for IEEE conference papers produced under \LaTeX\ using
%IEEEtran.cls version 1.7 and later.
%I wish you the best of success.


%\hfill January 11, 2007
\input{0_Abstract}
\input{1_Introduction}
\input{2_RelatedWorks}
\input{3_SystemModels}
\input{4_ProblemFormulationAndSolution}
\input{5_Simulation}
\input{6_Conclusion}
\bibliographystyle{abbrv}
\bibliography{Reference}

\end{document}

%% file: mymath_mod_JJS.tex
\newcommand{\node}[1]{\ensuremath{{\tt #1}}}

\newcommand{\separator}{
  \begin{center}
    \rule{\columnwidth}{0.3mm}
  \end{center}
}

\def\ie{{\it i.e.}}

\newcommand{\beq}{\begin{eqnarray*}}
\newcommand{\eeq}{\end{eqnarray*}}
\newcommand{\beqn}{\begin{eqnarray}}
\newcommand{\eeqn}{\end{eqnarray}}
\newcommand{\bemn}{\begin{multiline}}
\newcommand{\eemn}{\end{multiline}}

%% file: 0_Abstract.tex
\begin{abstract}
%\boldmath
% introduction
% issue : sensing coverage : delay from all area
% we study packet routing

Vehicular Sensor Network (VSN) is emerging as a new solution for monitoring 
urban environments such as Intelligent Transportation Systems and air pollution. 
One of the crucial factors that determine the service quality of 
urban monitoring applications is the delivery delay of sensing data packets in the VSN.
In this paper, we study
the problem of routing data packets with minimum delay in the VSN,
by exploiting
i) vehicle traffic statistics, ii) anycast routing and iii) knowledge of future trajectories of vehicles such as buses.
We first introduce a novel road network graph model that incorporates the three factors into the routing metric.
We then characterize the packet delay on each edge as a function of the vehicle density, speed and the length of the edge.
Based on the network model and delay function,
we formulate the packet routing problem as a Markov Decision Process (MDP)
and develop an optimal routing policy by solving the MDP.
Evaluations using real vehicle traces in a city show that our routing policy significantly improves
the delay performance compared to existing routing protocols.
%
%In this paper, we study a VSN routing problem to minimize
%the sensing packet delay from entire urban area,
%exploiting three representative features of VSN : (i) road statistics of moving vehicles,
%(ii) anycast network and (iii) knowledge of future trajectory from a certain type of vehicles such as a bus.
%In particular, we first build a novel road network graph which incorporates
%the effect of the above three VSN features. Then,
%characterize the packet delivery delay of each edge as a function of vehicle density, speed and the lengths of road segments. 
%Using the delay function, we formulate the packet routing problem as a Markov Decision Process (MDP)
%and develop an optimal routing policy.
%Evaluations using real vehicle traces in a city show that our optimal policy improves the delay performance by up to 70\%, compared to an existing routing protocol.

\end{abstract}

%Vehicular Sensor Network (VSN) is emerging as a new solution for monitoring 
%urban environments such as Intelligent Transportation Systems and air pollution. 
%One of the key problems in VSNs is to minimize the packet delivery delay, which 
%is crucial for the service quality of urban monitoring applications.
%[ Coverage story will be added ][ Using Vehicles' Trajectory Story (Using Bus) will be added ]

%In this paper, we study the problem of packet routing for anycast in VSNs. In particular, we characterize the packet delivery delay as a function of vehicle density, speed and the lengths of road segments. Using the delay function, we formulate the packet routing problem as a dynamic program, and develop a routing policy that minimizes the expected packet delivery delay.
%[ Bus Modeling ][ Data Coverage Increase ] 
%Evaluations using real vehicle traces in a city show that our optimal policy improves the delay performance by up to 70\%, compared to an existing routing protocol.

%% file: 1_Introduction.tex
%===================================================================
\section{Introduction} %\label{section_introduction}
%===================================================================
Recently, Vehicular Sensor Networks (VSNs) have received a great amount of attention as a new solution for monitoring the physical world\cite{MobEyes}. 
In VSNs, vehicles equipped with sensing devices move around an urban area and sense the urban environment periodically. 
The vehicles use vehicle to vehicle (V2V) or vehicle to infra (V2I) wireless communications to deliver the sensing data to an urban monitoring center.
%The VSN has mobile sensors (i.e., vehicles), and thus can monitor any area where vehicles can reach.
%Moreover, since sensor vehicles do not suffer from overheads such as deployment cost, processing limitations and power consumption,
%most of which are crucial to the traditional sensing systems with fixed sensor nodes,
%VSN can achieve sufficiently large sensing coverage with much lower cost.
Hence, unlike the traditional sensing system with fixed sensors that experiences limited coverage,
the vehicular sensing system can monitor any area where vehicles can reach.
Moreover, the vehicular sensor network can be deployed and maintained with relatively low cost
since it does not heavily rely on the network infrastructure for sensing data delivery.

%Moreover, sensor vehicles make VSN free to suffering from deployment overhead, power consumption and processing limits.
%However, in most of the traditional sensing systems,
%it is hard to achieve sufficient sensing coverage because of limited number or power of sensor nodes.}

%Their sensing data are transmitted over wireless channel of vehicle to vehicle (V2V) or vehicle to infra (V2I) to be delivered to a data monitoring center. 
%Compared to traditional sensing systems, VSNs can widely monitor the urban area by the vehicular mobility without suffering from deployment overhead, power consumption and processing limits. 
%Thus, VSNs have been studied for various applications such as urban traffic and environment monitoring and emergency information reporting.

Many of the VSN applications such as Intelligent Transportation System (ITS) require frequent updates of sensing information from all over the urban area, 
and hence it is important to guarantee \emph{timely delivery} of sensing data from \emph{every area} of interest to the urban monitoring center.
%Intelligent Transportation System (ITS) is one of the typical applications of VSN.
%It aims to provide innovative services relating to transport and traffic information for various users.
%The quality of ITS service is totally depending on the number and coverage of collected traffic data within an acceptable delay.
%Thus, for providing the high quality service in VSN application, 
%it is essential to efficiently route the sensed data to the wireless infrastructue 
%with fully exploiting V2V communication opportunities.
%
%This temporal and spatial 
Such a \emph{coverage guarantee} is rather challenging in VSNs
where the links (and thus the routes to destinations) can come and go depending on the mobility of vehicles. 
For instance, in such a network with intermittent connectivity, a vehicle sometimes has to carry the data while it moves away from the destination.
%packets are delivered to the denstination by carry-and-forward manners.
In fact, Delay-Tolerant Networks (DTNs) similarly experience intermittent routes to destinations, and there has been a large body of work that addresses the problem of routing data packets with minimum delay in DTNs\cite{VADD, MaxCon, TBD,MobiSpace,balasubramanian:dtnrouting}.
Due to the similarity, the packet routing policies for DTNs could be used for VSNs as well.
However, the VSN is distinguished from general DTNs in several aspects. 
First, vehicles in VSNs only move along the road, whereas mobile nodes in general DTNs are typically assumed to be able to move arbitrarily. 
Second, VSNs generally adopt anycast with multiple destinations, whereas most of the works in general DTNs assume unicast. 
%Third, in VSN, known future trajectories of some vehicles such as buses can be used for routing, 
Third, there are vehicles with predetermined future trajectories, such as buses,
whereas in general DTNs, %we cannot perfectly predict the future movements of the mobile nodes.
it is hard to predict the movement of mobile nodes.
Therefore, the packet routing policies for DTNs may not be directly applicable to VSNs, or may not be able to fully exploit the characteristics of VSNs. 
In this paper, we study the packet routing problem in the VSN with anycast.

\begin{figure}[]
  \centering
  \subfigure[Vehicle density on the road]{\epsfig{file=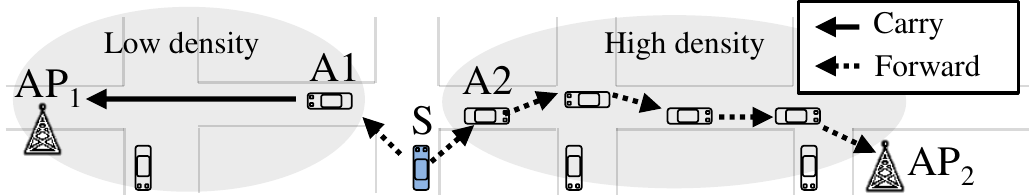,width=0.44\textwidth}\label{fig:intro1}}
  \subfigure[Anycast routing]{\epsfig{file=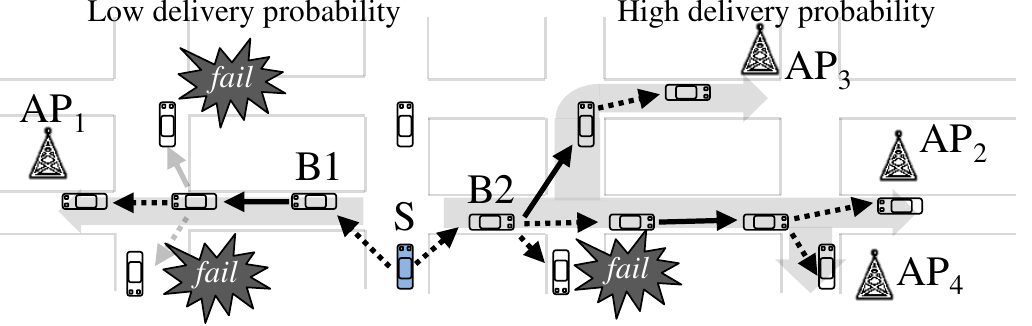,width=0.44\textwidth}\label{fig:intro2}}
  \subfigure[Knowledge of future trajectories]{\epsfig{file=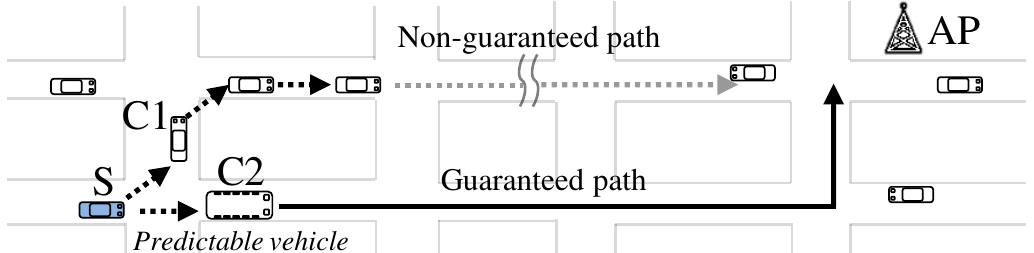,width=0.44\textwidth}\label{fig:intro3}}
  \caption{ Important factors on the delay performance in Vehicular Sensor Networks: APs are destinations in anycast routing
  \label{fig:intro}}
\end{figure}

In particular, we focus on minimizing the packet delivery delay from every area of interest to the urban monitoring center.
It is obvious that a packet routing algorithm with minimum delay must take into account the aforementioned characteristics of VSNs.
%under anycast in VSNs, with a focus on minimizing the packet delivery delay over all urban areas.
%To reduce data delivery delay in VSNs, the distinguished characteristics of VSNs should be considered to design routing algorithms.
%We propose a new routing algorithm which fully exploits the following three key features, as shown in Fig.~\ref{fig:intro}.
%\noindent{ \em 1) Density on routed path.}
First, since the vehicles can move only along the road, the vehicle density can be different from road to road.
Clearly, the road with high density can provide more opportunities of wireless multi-hop transfers,
and thus reduce the delivery delay on the road.
%high density road can provide more opportunities of 
%wireless multi-hop transfers, which in turn reduce the delivery delay on the road.
Consider a source vehicle $S$ in Fig.~\ref{fig:intro1}, which tries to select a better relay out of vehicles $A1$ and $A2$.
Even though $A1$ is closer to a destination (or AP in Fig.~\ref{fig:intro1}), forwarding to $A2$ may be more beneficial since 
the delay of multi-hop transfer over high density road is much smaller than carrying delay.
%Vehicles move through the road network with a regular movement tendency,
%hence we can estimate or obtain the various traffic statistics on roads, such as an amount of traffic and average speed of vehicles.
%Especially, a vehicle density represents the amount of wireless transmission opportunities on the roads.
%Fig.~\ref{fig:intro1} shows data packets in a higher vehicle density path are delivered faster from a source vehicle \node{S}
%to an AP by V2V transmissions. %
%\noindent{ \em 2) Number of destinations on routed path.}
Second, in anycast routing, a data packet just needs to be delivered to any one of the multiple destinations.
Hence, the effect of multiple APs can be exploited to reduce the packet delay.
%2) The topology of APs should be considered in terms of anycast routing in VSN.
As shown in Fig.~\ref{fig:intro2},
forwarding to $B1$ can fail to deliver packets to the targeted AP (\ie, $AP_1$) due to the uncertainty in $B1$'s movement.
However, since there exist many alternative APs on the direction of $B2$ (\ie, $AP_2$, $AP_3$ and $AP_4$),
forwarding to $B2$ may be a better option for reducing the delay.
%may succeed to deliver the packet to one of the alternative APs.
%Hence, $B2$ should be regarded as a better relay than $B1$.
%Due to anycast nature of VSNs, vehicles only need to deliver their packets to one of several destinations.
%However, due to the lack of connectivities, 
%it is not guaranteed that the packets are delivered to an originally intended destination.
%Thus, we should also consider alternative destinations for the case that a delivery to the intended destination is failed.
%Fig.~\ref{fig:intro2} shows that
%a routing path having a high delivery probability to one of any APs is better
%than a routing path having a minimum packet delay to be delivered one AP with low probability.
%\noindent{ \em 3) Known future trajectory.}
Third, the vehicles with known trajectories such as buses can help further reduce the delay.
%3) In VSNs, we can utilize the known future trajectories of certain type of vehicles (\ie, bus).
In Fig.~\ref{fig:intro3} where $S$ is far from the destination, such a predictable vehicle $C2$ 
guarantees to carry packets to the AP, which can significantly improve the routing performance compared to the delivery along a non-guaranteed path.
Note that the effect of known future trajectories is greatly appreciated in the scenario where the vehicle density is relatively low.

Our goal in this paper is to develop a delay-optimal packet routing algorithm in VSNs,
by taking into account the above ideas.
We first develop a novel road network graph model that incorporates the effect of predetermined vehicle trajectories
as well as unpredictable trajectories.
This network model is used to characterize the delay on a road segment as a function of the average vehicle density and speed,
and the length of the road segment.
Based on the network model and delay function,
we formulate the routing problem as a Markov Decision Process (MDP) that seeks to minimize the expected
delay of a packet from each area to one of the destinations. 
We develop an optimal packet routing algorithm by solving the MDP.
We examine our algorithm using Shanghai vehicle traces\cite{ShanghaiTaxi, BusTrace}, and show that the packet delay from each area
is significantly improved compared to existing routing algorithms in \cite{GPSR, DTN1_Vahdat}.

The rest of this paper is organized as follows:
In Section~\ref{section_2}, we discuss related work.
In Section~\ref{section_3}, we present the road network graph model of the VSN.
In Section~\ref{section_4}, we formulate the packet routing problem as an MDP, and develop
an optimal routing policy that solves the MDP.
In Section~\ref{section_5}, we evaluate the performance of our routing algorithm using real vehicle traces.
%
%Section~\ref{section_2} summarizes related works.
%Section~\ref{section_3} describes our system model of the VSN.
%Section~\ref{section_4} explains our packet routing problem formulation in the VSN.
%In Section~\ref{section_5}, we evaluate our optimal routing algorithm
%and conclude this paper in Section~\ref{section_6}.

%\subsection{Related Works}
%\subsection{Contributions and Organization}

%% file: 2_RelatedWorks.tex
%===================================================================
\section{Related Work}\label{section_2}
%===================================================================

% DTN 연구 / 차량 네트워크 연구 / DTN anycast 연구 / Trajectory 이용하는 연구 등등.. / Single Copy Multi Copy 등등
% 

%% 
There are a number of papers that study packet routing algorithms in VSNs\cite{MobEyes, dtnRoutingVSN, DAWN} and Vehicular Networks\cite{VADD, GPSR, MaxProp, MDDV, Trajectory, MaxCon, TBD, STDFS}.
Their common goal is to minimize the packet delivery delay to the destination.
The existing routing algorithms can be classified into multi-copy schemes and single-copy schemes.

In multi-copy routing schemes~\cite{MaxCon, SprayWait, MaxProp,balasubramanian:dtnrouting}, packets are replicated and forwarded
to have a better chance of reaching the destination.
However, such a replication can result in heavy congestion, which in turn hinders the packets from reaching the destination.
In single-copy routing schemes~\cite{VADD, TBD, STDFS, MobiSpace}, packets are not replicated,
but instead, the characteristics of vehicular networks are better utilized for reducing the packet delivery delay.
For instance, the Vehicle-Assisted Data Delivery (VADD) algorithm in \cite{VADD}
makes a routing decision based on the road layout, vehicle density and speed,
and is shown to outperform the routing algorithms that do not utilize the characteristics of vehicular networks.
The work in \cite{TBD} improves upon the VADD algorithm 
by taking into account predetermined trajectories of vehicles.
However, they do not fully exploit the knowledge of trajectories 
in that the entire trajectories and their impact on the delay performance are not incorporated into the routing metric.
All of these works assume unicast.
In this paper, we investigate the routing problem in VSNs by formulating a Markov Decision Process (MDP) 
that fully takes into account the effect of predetermined future trajectories and anycast routing as well as the vehicle density and speed.

%% file: 3_SystemModels.tex
\section{System Model}\label{section_3}
Our vehicular sensor network is modeled as a ``vehicular sensing system" working on an urban area or a ``road network" described in the following.
%In this section, we describe the vehicular sensing systems, and explain our assumptions
%to model the vehicular sensing systems. 
%We model the roadmap as the directed graph, 
%and suggest our routing scheme over the directed graph.
%Also, we modify the directed graph by adding links came from the predetermined path of the buses,
%and suggest the way to use the addtional links for our routing.

%=================== Subsection 1 ======================
\subsection{Vehicular Sensing System}

We consider a Vehicular Sensor Network (VSN) that consists of vehicles and WiFi Access Points (APs).
Vehicles moving along the road sense the urban area, generate sensing data packets periodically,
and deliver the packets to one of the APs by carrying or forwarding to others.
The APs are deployed only at intersections, and connected to the urban monitoring center via wired backhaul networks.
Hence, the sensing data packets just need to be delivered to one of the APs.
There are two types of vehicles including those with predetermined trajectories (such as buses and police patrol vehicles) and those with unpredictable trajectories (such as taxis and cars).
For simplicity of exposition, a vehicle with predetermined trajectory will be called ``bus" throughout the paper.
As in a real city, we assume that a certain fraction of vehicles in the VSN are buses (i.e., vehicles with predetermined paths).

We assume that vehicles can use the digital roadmap and their GPS information, and are equipped with the IEEE 802.11 devices to communicate with other vehicles or the APs.
We also assume that once a vehicle forwards a packet to another vehicle, the packet is immediately deleted from the sender vehicle; so that there is always at most one copy of each data packet in the network.
\subsection{Road Network Graph}

The urban area or road network to be sensed by vehicles is modeled as a graph $G=(\mathcal{I},\mathcal{R})$ where $\mathcal{I}$ 
is the set of intersections and $\mathcal{R}$ is the set of road segments connecting the intersections.
The network $G$ is a directed graph, and hence, road segment $e_{ij} \in R$ denotes the road from intersection $i$ to (neighboring) intersection $j$.
Denote by $\mathcal{I_{AP}} \subset \mathcal{I}$ the set of intersections where APs are placed.
In our system, there are $N$ vehicles in total, and 
we define $\mathcal{V} = \{ \begin{matrix}0,1, \ldots, M \end{matrix} \}$ as the set of the types of vehicles. 
If the type $v \in \mathcal{V}$ of a vehicle is zero, its trajectories are unpredictable.
Otherwise, if $v \neq 0$, then it represents a bus line with a predetermined route.
Thus, $M$ is the number of bus lines in the VSN.

\begin{figure}[t!]
  \centering
  \subfigure[Roadmap]{\epsfig{file=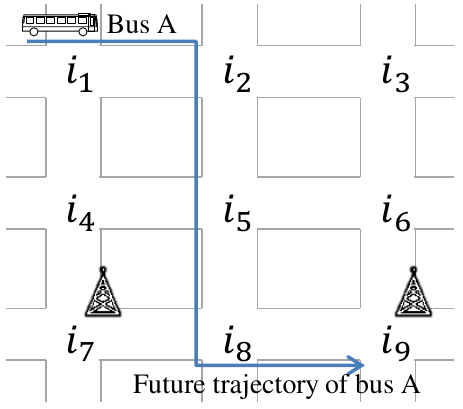,width=0.24\textwidth}\label{fig:road networks a}}
  \subfigure[Road network graph $G$]{\epsfig{file=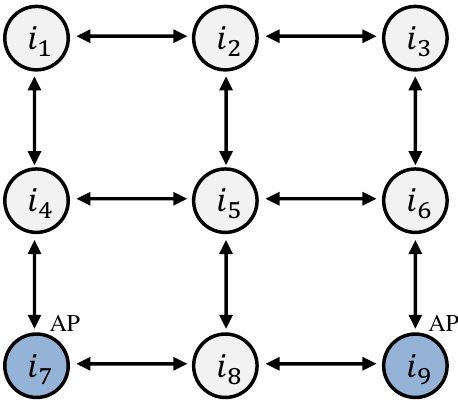,width=0.24\textwidth}\label{fig:road networks b}}
  \caption{Roadmap and its corresponding road network graph
  \label{fig:road networks}}
\end{figure}
% (preferably heading toward the destinations). 
%\begin{figure*}[t! ht]
%  \centering
%  \subfigure[Roadmap]{\epsfig{file=figures/figure_graph_1.pdf,width=0.24\textwidth}\label{fig:road networks a}}
%  \subfigure[Road network graph $G$]{\epsfig{file=figures/figure_graph_2.pdf,width=0.24\textwidth}\label{fig:road networks b}}
%\subfigure[Road network graph $G$]{\epsfig{file=figures/figure_graph_3.pdf,width=0.44\textwidth}\label{fig:road networks b}}
%  \caption{Roadmap and its corresponding road network graph
%  \label{fig:road networks}}
%\end{figure*}

Fig.~\ref{fig:road networks} shows the roadmap and its corresponding directed graph $G$.
In Fig.~\ref{fig:road networks a}, two APs are placed at the intersections $i_7$, $i_9$
and the path of bus \node{A} is the sequence of intersections, $i_1$, $i_2$, $i_5$, $i_8$ and $i_9$.
Note that the path of a packet is a sequence of consecutive road segments and intersections
since the packets are carried and forwarded by vehicles moving along the road.
Unlike the usual communication network where there is a fixed set of routes all the time,
in the VSN, the links on a ``data path" are formed by the mobility of vehicles,
and thus, they do not always exist.
Accordingly, the road network graph $G$ represents a network that can be ``potentially" used for delivering data packets,
and the existence of data links in the network is highly uncertain.
Therefore, the data packets are delivered as if they are routed over a random graph,
and this will be accounted for in our formulation of the packet routing problem in Section~\ref{section_4}.

\begin{figure}[!h]
  \centering
  \epsfig{file=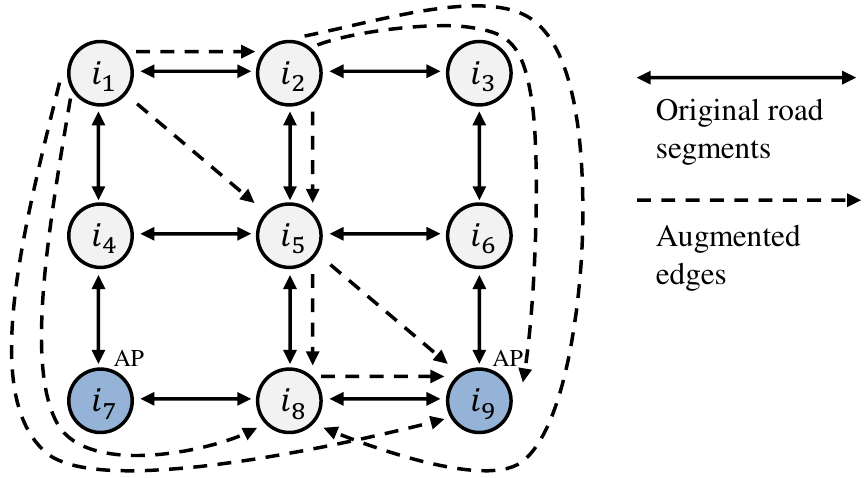,width=0.42\textwidth}
  \caption{Augmented road network graph $G'$ incorporating bus line A
   \label{fig:road network graph with bus}}
\end{figure}

To incorporate the effect of buses into the graph,
we note that a bus can carry its packets not only to the neighbor intersections but 
also to every intersection along its future trajectory with 100\% probability in a certain time.
Hence, it is as if there is an edge directly connecting an intersection to another intersection which is multiple blocks away.
%we can add the edges created by the buses from each intersections to other intersections.
%
To define these additional edges,
we introduce a new notation $e_{ij}^v$ representing the edge from $i$ to $j$ created by type-$v$ vehicle.
%of a type of vehicles $v \in \mathcal{V}$ into the edges \ie, $e_{ij}^v$.
%Thus, we make a new defition of edges as $e_{ij}^v$ where $v \in \mathcal{V}$ and $i, j \in \mathcal{I}$.
We denote the set of newly added edges by $\mathcal{L}$
and a new road network graph by $G'=(\mathcal{I},\mathcal{R}')$ where $\mathcal{R}' = \mathcal{R} \cup \mathcal{L}$.
Note that edge $e_{ij}^0 \in \mathcal{R}'$ is the same as $e_{ij} \in \mathcal{R}$ of graph $G$.
Let $\mathcal{R}'_s$ be the set of edges in $\mathcal{R}'$ corresponding to a ``single" road segment in $\mathcal{R}$,
\ie, $\mathcal{R}'_s=\{ e^v_{ij} \in \mathcal{R}': \exists e_{ij} \in \mathcal{R} \}$.
Fig.~\ref{fig:road network graph with bus} shows the new road graph $G'$, 
which is augmented from $G$ in Fig.~\ref{fig:road networks b} to take into account the effect of bus \node{A}'s predetermined path.
% the pre-determined paths of the bus \node{A} as well as
%probabilistic path of graph $G$ in Fig.~\ref{fig:road networks b}.
Using the graph $G'$, we formulate the delay-optimal packet routing problem in Section~\ref{section_4}.

%% file: 4_ProblemFormulationAndSolution.tex
\section{Delay-Optimal Routing Algorithm} \label{section_4}
%===================================================================
In this section,
we develop a routing policy that minimizes the packet delay to any one of the APs.
In particular, we formulate the packet routing problem as a Markov Decision Process (MDP)
and find an optimal routing policy that solves the MDP.
We also discuss how vehicle traffic statistics can be used to estimate the parameters in the MDP
such as state transition probability and (link delay) cost.
%
%In this section, 
%we first introduce a concept of our routing algorithm,
%which makes routing decisions in each intersection.
%From the Markov Decision Process (MDP) formulation, 
%we find the optimal routing decision which minimizes the expected packet forwarding delay
%for every areas in networks. 
%Also, we explain the detailed formulation of the data forwarding probability function and 
%the expected data forwarding delay through the link in a graph $G'$ used in our MDP formuation.
% MDP를 
%================================== Routing Algorithm Overview ===================================
\subsection{Routing Algorithm Overview}\label{subsection4_overview}
As mentioned in Section~\ref{section_3},
packets are delivered by vehicles along the intersections and edges in the augmented road network graph $G'$.
We assume that the routing policy is computed in advance using the vehicle traffic statistics,
and the vehicles only have a routing table that can be used for forwarding packets.
This would reduce the amount of online computations and thus enable fast forwarding of packets.
Our routing algorithm specifies the forwarding decision at every intersection and edge as follows:

\subsubsection{At intersections}
Consider a vehicle arriving at an intersection, and assume that it has data packets.
Clearly, the vehicle can forward its packets to a neighbor intersection if it meets another vehicle heading to
the neighbor intersection or if it moves to the neighbor intersection.
As mentioned above, a precomputed routing policy is loaded on the vehicles,
and hence, a routing policy that specifies only a single best next hop does not work in our setting.
Note that if the edge selected as a single best next hop is not available due to no vehicles moving along the edge,
such a routing cannot continue simply because it does not
define what to do in that case.
Consequently, a contingency decision-making framework is necessary.

Our idea is to prioritize the outgoing edges of each intersection.
Thus, if the vehicle does not either meet another vehicle along the edge with the first priority or move onto the edge, 
then it attempts packet forwarding toward the edge with the second priority, and so on.
In Section~\ref{subsection4_problem_formulation},
we develop a prioritization method (\ie, routing policy) that minimizes the packet delay.

\subsubsection{On edges}
The packet forwarding on an edge, say $e_{ij}^v$, is divided into two cases.
If $j$ is a neighbor intersection of $i$ in the original network graph $G$ (\ie, $e_{ij}^v \in \mathcal{R}'_s$),
then a vehicle on $e_{ij}^v$ forwards its packets to a vehicle closer to $j$.
If $j$ is not a neighbor intersection of $i$ in $G$ (\ie, $e_{ij}^v \in \mathcal{R'} \setminus \mathcal{R}'_s$), 
then $e_{ij}^v$ is an augmented edge by the bus with type $v$.
On those edges, the bus with the corresponding type carries packets to $j$.

%===================================== Problem Formulation ======================================
\subsection{MDP Formulation and Optimal Routing Policy} \label{subsection4_problem_formulation}
Markov Decision Process (MDP) provides a mathematical framework for modeling the decision-making
system where the transition from a state to another state is probabilistic, 
and the transition probability depends on the decision taken at the state.
Hence, MDP can effectively capture the essence of the routing problem in VSNs
where the delivery of a packet from an intersection to another is subject to uncertainty as there
may not always exist a vehicle moving in the desired direction.
%
%To find the optimal routing policy that minimizes the expected data delivery delay at each intersection, 
%We formulate the routing problem as a Markov Decision Process (MDP)[CITE].
%The MDP provides a mathematical framework to model the decision making
%in case where transitions between each state are partly random and partly under the control of decision maker.
%%{\bf JS: A simple description about MDP goes here to be more understandable for mapping the VSN scenario to the states, controls in MDP. }
%
The set of states in our MDP represents the set of intersections $\mathcal{I}$,
and the transitions from one state to others occur probabilistically over the edges $e_{ij}^v$ in $\mathcal{R'}$.
Then, the control decision at each state in the MDP corresponds to a routing decision at each intersection.
Note that the state transition probability from state $i$ to $j$ 
depends on the vehicle traffic statistics and the routing decision at intersection $i$.
Denote by $\bm u_i$ a routing decision at intersection $i$.
To account for the prioritization discussed above, 
$\bm u_i$ is defined as a row vector, $\bm u_i = [ \begin{matrix} u^1_{i} & u^2_{i} & \ldots &u^{K_{i}}_{i} \end{matrix} ]$, where $u^1_{i}, u^2_{i}, \ldots ,u^{K_{i}}_{i} \in \mathcal{R'}$ are all the outgoing edges from intersection $i$ and $K_i$ is the total number of outgoing edges from $i$.
The order of elements in $\bm u_i$ represents the priority, that is, $u_i^k$ indicates
the $k$-th most preferred next hop from intersection $i$ for minimum packet delay.
Let $\mathcal{U}(i)$  be the set of all possible decisions $\bm u_i$ at intersection $i$,
then the size of $\mathcal{U}(i)$ is given by $|\mathcal{U}(i)|=K_i!$.
%Clearly, there are $K_i!$ possible forwarding policies, \ie, $|\mathcal{U}(i)| = K_i!$.
%Denote by $\mathcal{O}(i)$ the set of the outgoing edges from the intersection $i$.

As mentioned above, the routing decision $\bm u_i$ affects the data forwarding probability from intersection $i$ to other intersections.
Let $P^{v}_{ij}(\bm u_i)$ be the probability that a packet is forwarded from intersection $i$ to $j$ by a type-$v$ vehicle under a routing decision $\bm u_i$.
%Modeling the data forwarding probability as a function of a routing decision 
%enable us to formulate anycast routings problems.
%We will explain the details in the section~\ref{subsection4_data_probability}.
%Note that the works~\cite{VADD,TBD} fix the data forwarding probability 
%based on the position of the destination, so that their works cannot be directly applied to 
%decide which destination is the best to minimize the packet delay in anycast scenarios.
Denote by $d^v_{ij}$ the expected data delay on an edge $e_{ij}^v \in \mathcal{L}$
which is the time it takes to carry and forward a packet along the edge $e_{ij}^v$.
The delay $d_{ij}^v$ can be estimated using the average vehicle speed, density and the total length of road segments from intersection $i$ to $j$.
We discuss the details of the estimation of $d^v_{ij}$ in Section~\ref{subsection4_link_delay}.

\begin{figure}[]
\centering
\subfigure[Outgoing edges at intersection $i$]{\epsfig{file=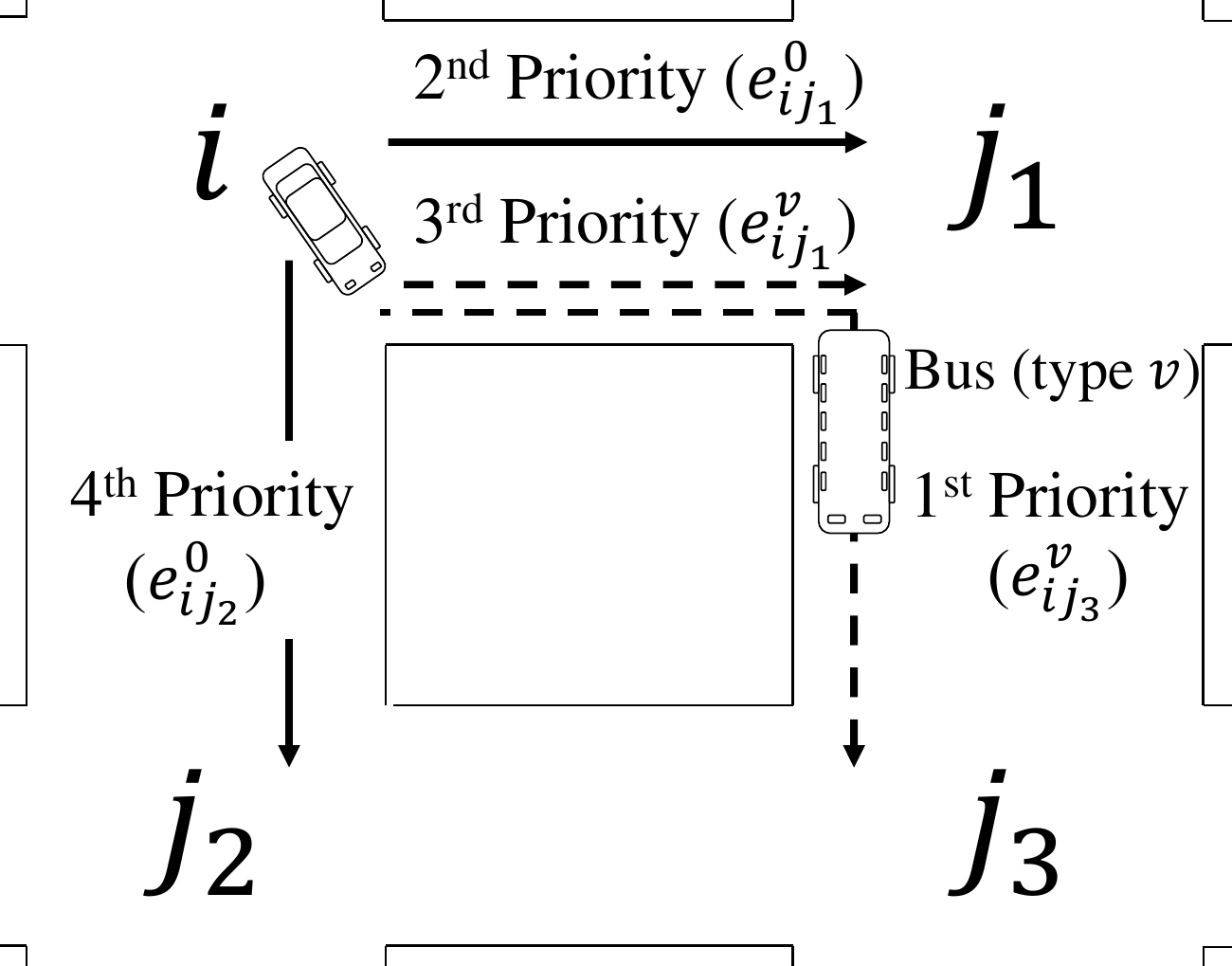,width=0.24\textwidth}\label{fig:MDP a}}
\subfigure[Markov decision process (MDP) model of (a)]{\epsfig{file=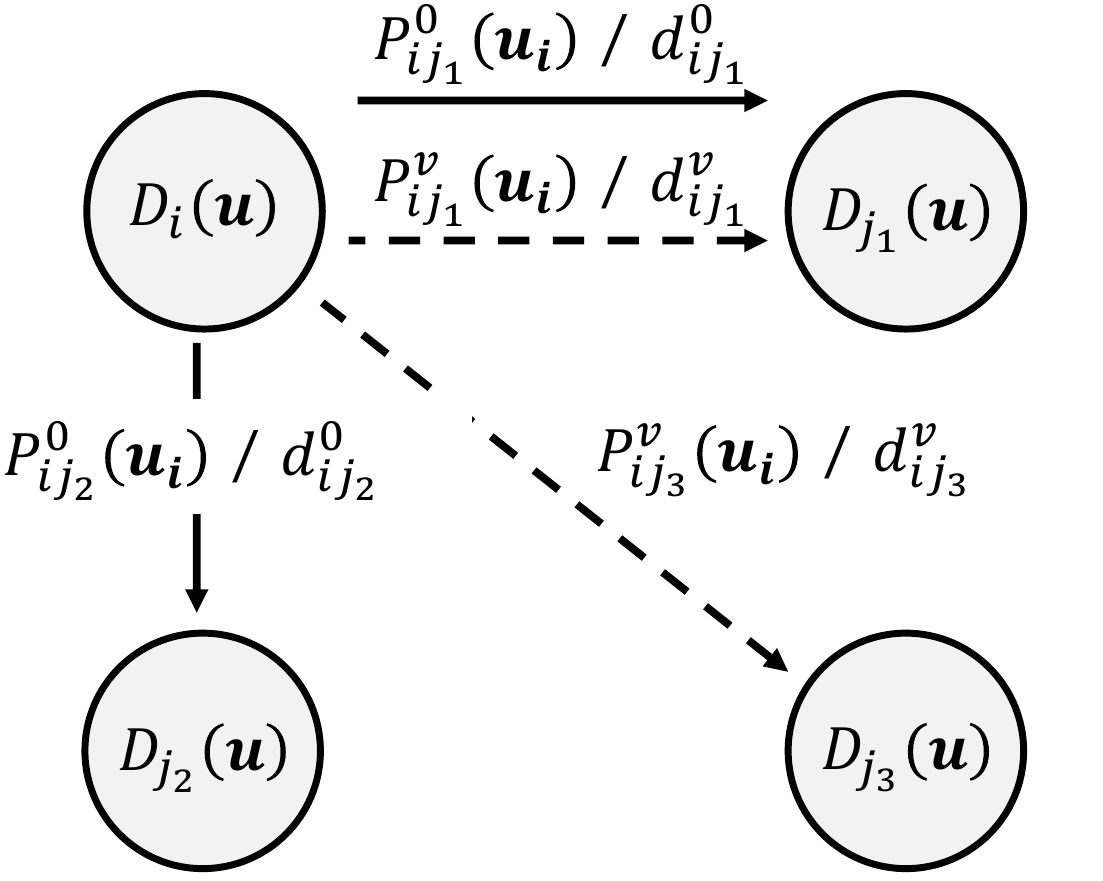,width=0.24\textwidth}\label{fig:MDP b}}
\caption{Formulation of routing problem in graph $G'$
\label{fig:MDP}}
\end{figure}

Under a routing policy $\bm u =[ \bm u_i, \forall i \in \mathcal{I}]$,
let $D_i(\bm u)$ be the expected data delivery delay from intersection $i$ to any AP.
Thus, for an intersection $i$ where an AP is placed (\ie, $i \in \mathcal{I_{AP}}$),
$D_i(\bm u)=0$ for any routing policy $\bm u$ and
no action is taken at $i$.
On the other hand, at an intersection $i$ where there is no AP, 
$D_i(\bm u)$ depends on $d^v_{ij}$, $D_j(\bm u)$ and $P^{v}_{ij}(\bm u_i)$ for every outgoing edge $e_{ij}^v$ from $i$.
For better understanding, in Fig.~\ref{fig:MDP}, we illustrate an example of the routing decision $\bm u_i$ and
related parameters.
Assume that there exist four possible forwarding candidates which are prioritized by $\bm u_i$ as in Fig.~\ref{fig:MDP a}.
Based on the routing scenario, the MDP model in Fig.~\ref{fig:MDP b} has four outgoing edges which correspond to the forwarding candidates,
and specifies the forwarding probability $P^{v}_{ij}(\bm u_i)$ and the edge delay $d^v_{ij}$.
Clearly, $D_i(\bm u)$ can be computed as 
\begin{align}
D_i(\bm u) =& P_{ij_1}^0 \times \left(d_{ij_1}^0 + D_{j_1}(\bm u)\right) +  P_{ij_2}^0 \times \left(d_{ij_2}^0 + D_{j_2}(\bm u)\right) \nonumber
\\
& + P_{ij_1}^v \times \left(d_{ij_1}^v + D_{j_1}(\bm u)\right) +  P_{ij_2}^v \times \left(d_{ij_2}^v + D_{j_2}(\bm u)\right)
\end{align}
In general, $D_{i}(\bm u)$ can be expressed as follows:
\begin{equation} \label{eq:value}
D_i(\bm u)  = 
\sum_{v \in \mathcal{V}} \sum_{j \in \mathcal{I}}P_{ij}^v( \bm u_i) \cdot \left( d^v_{ij} + D_j(\bm u) \right)
\text{, $i \in \mathcal{I} \setminus \mathcal{I_{AP}}$.}
\end{equation}
Hence, our routing problem can be formulated as 
\begin{equation}
\min_{\bm u} D_i(\bm u), \forall i.
\end{equation}
%

%<<<<<<< .mine

The optimal solution $\bm u^*$ to the above problem gives a routing policy that minimizes the expected delay from $i$ to any one of the APs.
The routing problem can be solved using the value iteration method\cite{DP} (see Algorithm~\ref{value_iteration}).
For a given initial delay vector $\bm D^0$, the expected delay from intersection $i$ is updated as in \eqref{eq:value_iteration}.
The iteration is terminated if the two consecutive delay vectors $\bm D^{k}$ and $\bm D^{k-1}$
are close enough, \ie,
\begin{equation}
\max_{i \in \mathcal{I}} |D_i^{k}-D_i^{k-1}| < \epsilon
\end{equation}
where $\epsilon$ is a predetermined threshold value.
It is known that for each $i$ the sequence $\{ D_i^k \}$ generated by the iteration in~\eqref{eq:value_iteration}
converges close to its optimal value $D^*_i = D_i(\bm u^*)$ after a sufficient number of iterations\cite{Bellman}.
The optimal routing policy $\bm u^* =[ \bm u_i^*, \forall i \in \mathcal{I}]$
is then computed using the estimated optimal delay vector $\bm D^k = [ D_{i}^k, \forall i \in \mathcal{I} ]$,
as in~\eqref{eq:optimal_policy}.
%
%directly provides the optimal routing policy at each $i$
%that minimizes the expected delay from $i$ to any one of the APs.
%Borrowing value iteration method\cite{Bellman}, we solve the MDP by iteratively updating $D_i$ in (\ref{eq:value}) for all $i \in \mathcal{I}$,
%We introduce the routing policy computation algorithm~\ref{value_iteration} as follows.
%=======
\vspace{-0.2cm}
%>>>>>>> .r137
\begin{algorithm}
\caption{Routing Policy Computation  \label{value_iteration}}
  \begin{compactenum}
 \item {\bf Procedure} ComputingOptimalPolicy($\bm D^0$)
 \item \hspace{0.5cm} {\bf Input:} initial value $\bm D^0=[D^0_i, \forall i \in \mathcal{I}]$
 \item \hspace{0.5cm} {\bf Output:} optimal routing policy $\bm u^*= [\bm u^*_i, \forall i \in \mathcal{I}]$
 \item \hspace{0.5cm} {\bf Local variable:} $k = 0$
%\item  \hspace{1.2cm} $D^{k+1}_i = \min \limits_{\bm u_i \in \mathcal{U}(i)}\sum \limits_{ v \in \mathcal{V}}\sum \limits_{j \in \mathcal{I} } P_{ij}^v(\bm u_i) \cdot (d_{ij}^v+D^{k}_j)$
 \item \hspace{0.6cm} {\bf repeat} 
% \item \hspace{1.2cm} $\bm D \leftarrow \bm D'$
% \item \hspace{1.2cm} {\bf for each} state i {\bf do}
\vspace{+0.2cm}\item \hspace{1.2cm} 
%\vspace{+0.4cm}\item \hspace{1.2cm} 
%\vspace{-0.67cm}
\vspace{-0.69cm}
\begin{equation} \hspace{1.3cm} D^{k+1}_i = \min \limits_{\bm u_i \in \mathcal{U}(i)}\sum \limits_{ v \in \mathcal{V}}\sum \limits_{j \in \mathcal{I} } P_{ij}^v(\bm u_i) \cdot (d_{ij}^v+D^{k}_j)
\label{eq:value_iteration} \end{equation}

\vspace{-0.1cm}
 \item \hspace{1.2cm} $k=k+1$
% \item \hspace{1.2cm} {\bf end} 
 \vspace{+0.2cm} \item \hspace{0.5cm} {\bf until} $\max\limits_{i \in \mathcal{I}}|D^k_i-D_i^{k-1}| < \epsilon $
% \item \hspace{0.6cm} 
\vspace{+0.4cm}\item \hspace{1.2cm} 
\vspace{-0.67cm}
\begin{equation}\hspace{0.6cm} \bm u_i^* = \argmin\limits_{\bm u_i \in \mathcal{U}(i)} \sum \limits_{ v \in \mathcal{V}}\sum \limits_{j \in \mathcal{I} } P_{ij}^v( \bm u_i) \cdot \left( d^v_{ij} + D^k_j \right) \text{, } \forall i \in \mathcal{I}
\label{eq:optimal_policy} 
\end{equation}
 \item {\bf return} $\bm u^*$
\end{compactenum}
\end{algorithm}

\textbf{Remark:}
In our anycast setting, multiple APs are deployed at intersections,
and each intersection $i$ with an AP will have $D_i(\bm u)=0$.
Thus, the optimal routing policy solving the MDP would try to forward the packets
toward one of the intersections with APs.
Therefore, our routing policy can take advantage of multiple destinations in anycast routing.

%
%After taking an initial value $\bm D^0$, for each intersection $i \in \mathcal{I}$,
%the expected delay in the next step (\ie, $D_i'$) is calculated by the expected delays of neighboring 
%intersections in the present step (\ie, $D_j$)
%until the expected delays calculated on two successive steps are close enough, \ie,
%\begin{equation}
%\max_{i \in \mathcal{I}}|D_i-D_i'| < \epsilon
%\end{equation}
%where $\epsilon$ is a predetermined threshold value.
%%
%For a proper initializations, value iteration guarantees that the expected delay converges close to its optimal value after a sufficient number of iterations\cite{Bellman}.
%From the result of the algorithm~\ref{value_iteration},
%the optimal routing policy $\bm u^* =[ \bm u_i^*, \forall i \in \mathcal{I}]$,
%which minimize the expected pacekt delivery delay $D_i(\bm u)$ for every intersections (\ie, $\forall i \in \mathcal{I}$),
%can be calculated. 
%as follows:
%%
%\begin{equation} \label{eq:optimal_policy}
%\bm u_i^* = \argmin_{ \bm u_i \in \mathcal{U}(i)} \sum_{ v \in \mathcal{V}}\sum_{j \in \mathcal{I} } P_{ij}^v( \bm u_i) \cdot \left( d^v_{ij} + D_j \right)
%\end{equation}

%===================================== Data Fowarding Probability ======================================

\input{4_2_Probability}
\input{4_3_LinkCost}

%% file: 4_2_Probability.tex
%===================================================================
\subsection{Data Forwarding Probability $P_{ij}^v(\bm u_i)$} \label{subsection4_data_probability}
\begin{figure}[!h]
  \centering
  \epsfig{file=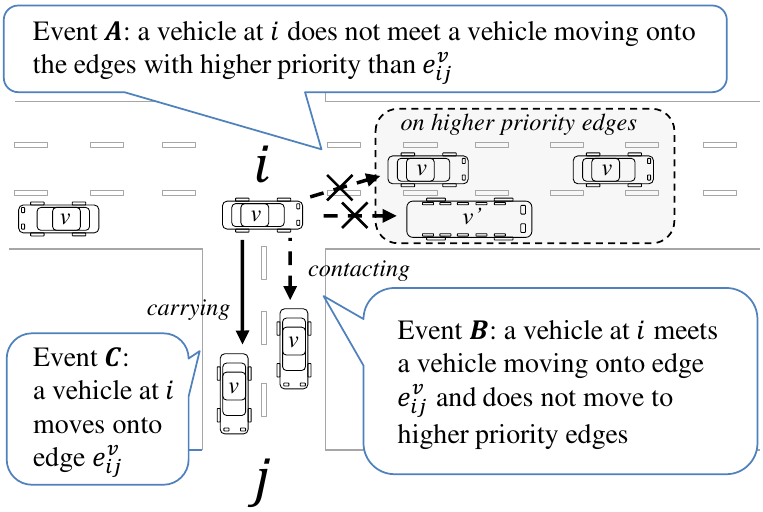,width=0.44\textwidth}
  \caption{Conditions for the forwarding of a packet from $i$ to $j$.
   \label{fig:probability}}
\end{figure}
Next, we discuss how to calculate the data forwarding probability $P^v_{ij}(\bm u_i)$.
Let $\mathcal{O}(i)$ be the set of outgoing edges from intersection $i$.
At an intersection $i$ where there is no AP,
a vehicle forwards or carries packets to a neighbor intersection along one of the edges in $\mathcal{O}(i)$.
Thus, $P^v_{ij}(\bm u_i)$ is a function of the probabilities $Q_{ij}^v$ and $C_{ij}^v$  defined as
\begin{itemize}
\item[--] $Q_{ij}^v$: probability that a vehicle at $i$ moves onto edge $e_{ij}^v$
\item[--] $C_{ij}^v$: probability of contacting a vehicle moving onto $e_{ij}^v$.
\end{itemize}
First, we find an expression for $P^v_{ij}(\bm u_i)$ in terms of $Q_{ij}^v$ and $C_{ij}^v$,
and then describe how to estimate $Q_{ij}^v$ and $C_{ij}^v$ using vehicular traffic statistics.
This will provide a complete description of the computation of $P^v_{ij}(\bm u_i)$.

\subsubsection{Computation of $P^v_{ij}(\bm u_i)$}
Consider an event that packets at intersection $i$ are forwarded to $j$ through an edge $e_{ij}^v$ under a routing decision $\bm u_i$.
Clearly, this forwarding event can occur if a vehicle with the packets at $i$ meets another vehicle moving onto $e_{ij}^v$ or
it moves onto $e_{ij}^v$.
The additional condition for the forwarding event to occur is that 
a vehicle at $i$ does not encounter vehicles moving onto the edges with higher priority than $e_{ij}^v$ in $\bm u_i$ and 
it does not move onto those edges.
%to occur the event under a routing policy $\bm u_i$,
%it is also satisfied that the vehicle at $i$ does not meet vehicles moving to edges with higher priority than $e_{ij}^v$ and also
%not move to those edges by itself.
Those conditions are illustrated by three events in Fig.~\ref{fig:probability} that are defined as
\begin{itemize}
\item[--] $\bm A$: the event that a vehicle at $i$ does not meet a vehicle moving onto the edges with higher priority than edge $e_{ij}^v$
\item[--] $\bm B$: the event that a vehicle at $i$ meets another vehicle moving onto $e_{ij}^v$ and it does not move onto the edges with higher priority than $e_{ij}^v$
\item[--] $\bm C$: the event that a vehicle moves onto $e_{ij}^v$.
\end{itemize}
Then $P^v_{ij}(\bm u_i)$ can be expressed as
%
%and let $Pr(\bm A)$ be the probability of the occurrence of event \textbf{A}.
%Event \textbf{B} represents that the vehicle at $i$ meets a vehicle moving to $e_{ij}^v$ and it 
%does not move to higher priority edges than $e_{ij}^v$, and event \textbf{C} is that the vehicle moves to $e_{ij}^v$.
%Also, we denote $Pr(\bm B)$ and $Pr(\bm C)$ as the probability of the occurrence of event \textbf{B} and \textbf{C}, respectively.
%Then, $P^v_{ij}(\bm u_i)$ is described as follows:
\begin{align}
P_{ij}^{v}(\bm u_i) &=  Pr\bigl[\bm A \cap (\bm B \cup \bm C)\bigr] \nonumber \\ 
& = Pr(\bm A) \times Pr(\bm B \cup \bm C)
\label{eq:Pr}
\end{align}
where $Pr(\bm E)$ is the probability of event $\bm E$.
The equality follows from the fact that the moving direction of a vehicle is independent of that of others.
Using $Q_{ij}^{v}$ and $C_{ij}^{v}$, 
\eqref{eq:Pr} can be rewritten as follows:
\begin{align}
P_{ij}^{v}(\bm u_i) &= Pr(\bm A) \times \left[ Pr(\bm B) + Pr(\bm C) - Pr(\bm B \cap \bm C) \right] \nonumber \\
&= Pr(\bm A) \times \left[ Pr(\bm B) + Pr(\bm C) - Pr(\bm B| \bm C)Pr(\bm C) \right] \nonumber \\
&=\left[\prod_{ \scriptscriptstyle{e_{ik}^{w} \in}
\mathcal{H}(\bm u_i,e_{ij}^v)}(1-C^{w}_{ik})\right] \nonumber
\\ 
&\hspace{0.3cm}\times \left[C_{ij}^{v}(1-\sum \limits_{\scriptscriptstyle{e_{ik}^{w} \in} \mathcal{H}(\bm u_i,e_{ij}^v)}Q_{ik}^{w})+Q_{ij}^v-C_{ij}^vQ_{ij}^v\right]
\label{eq:P}
\end{align}
%
%\begin{align}
%P_{ij}^{v}(\bm u_i) &= Pr(\bm A) \times \left[ Pr(\bm B) + Pr(\bm C) - Pr(\bm B| \bm C)Pr(\bm C) \right] \nonumber \\
%&= 
%&=\left[\prod_{ \scriptscriptstyle{e_{ik}^{w} \in}
%\mathcal{H}(\bm u_i,e_{ij}^v)}(1-C^{w}_{ik})\right] \nonumber
%\\ 
%&\times \left[C_{ij}^{v}(1-\sum \limits_{\scriptscriptstyle{e_{ik}^{w} \in} \mathcal{H}(\bm u_i,e_{ij}^v)}Q_{ik}^{w})+(1-C_{ij}^v)Q_{ij}^v\right]
%\label{eq:P}
%\end{align}
%
where $\mathcal{H}(\bm u_i,e_{ij}^v)$ is the set of the edges which have higher priority than $e_{ij}^v$ in a routing decision $\bm u_i$.
The first product term in \eqref{eq:P} corresponds to $Pr(\bm A)$, \ie, the probability that a vehicle at $i$ does not meet a vehicle moving onto
higher priority edges than the edge $e_{ij}^v$ in routing decision $\bm u_i$.
The second product term is equal to $Pr(\bm B \cup \bm C)$, \ie, 
the probability that a vehicle at $i$ carries or forwards its packets onto edge $e_{ij}^v$.

%Similar to \eqref{eq:Pr}, we can calculate $P^v_{ij}(\bm u_i)$ by using $Q_{ij}^{v}$ and $C_{ij}^{v}$ under $\bm u_i$.
%We define $\mathcal{H}(\bm u_i,e_{ij}^v)$ as a set of the edges which have higher priority than $e_{ij}^v$ in a routing decision $\bm u_i$, 
%and then $P^v_{ij}(\bm u_i)$ can be calculated as~\eqref{eq:P}.
%The first product term of \eqref{eq:P} means
%the probability that a vehicle at $i$ does not meet vehicles which move through higher priority edges than the edge $e_{ij}^v$.
%The second product term means the summation of two mutually exclusive probabilities that 
%1) a vehicle at $i$ contacts the vehicle moving through $e_{ij}^v$ and does not move to the higher priority edges
%2) a vehicle at $i$ does not contact the vehicle moving through $e_{ij}^v$ and moves to $e_{ij}^v$.

\subsubsection{Estimation of $Q_{ij}^v$ and $C_{ij}^v$}
Recall that $Q_{ij}^v$ is the probability that a vehicle at intersection $i$ moves onto edge $e_{ij}^v$,
and $C_{ij}^v$ is the probability of contacting a vehicle moving onto $e_{ij}^v$.
Obviously, $Q_{ij}^v$ and $C_{ij}^v$ are determined by the parameters such as the vehicle density
and moving tendency.
In particular, the following parameters are used to express $Q_{ij}^v$ and $C_{ij}^v$:
\begin{itemize}
\item[--] $q_{ij}^0$: the fraction of type-$0$ vehicles moving to a neighbor intersection $j$ among all vehicles which arrive to $i$.
\item[--] $p_{ij}^0$: the probability of meeting a type-$0$ vehicle at $i$ that moves to $j$.
\item[--] $q_{i}^v$: the fraction of type-$v$ vehicles among all vehicles which arrive to $i$,
\item[--] $p_{i}^v$: the probability of meeting a type-$v$ vehicle at $i$.
\end{itemize}
Note that these parameters can be extracted from vehicle traffic statistics \cite{VADD}.
%
%Recall that one of our assumptions is that we can easily get vehicular traffic statistics from the traffic data center.
%Thus, we can obtain vehicles' moving and contacting probabilities at intersection $i$,
%which can be easily estimated from the vehicles' moving tendency at $i$ as described in \cite{VADD}.
%Denote $q_{ij}^0$ as the fraction of type-$0$ vehicles moving to a neighbor intersection $j$ among all type-$0$ vehicles which arrive to $i$,
%and $p_{ij}^0$ as the probability of meeting a type-$0$ vehicle at intersection $i$ that moves to $j$.
%Let $q_{i}^v$ be the fraction of type-$v$ vehicles among all vehicles which arrive to $i$,
%and $p_{i}^v$ the probability of meeting a type-$v$ vehicle at $i$.

\begin{figure}[]
  \centering
  \epsfig{file=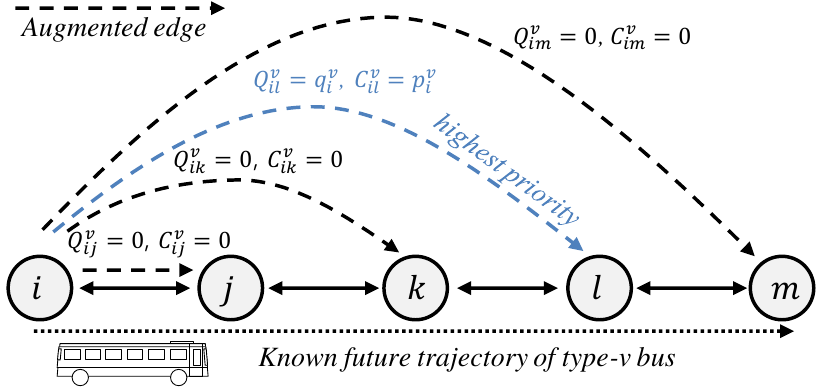,width=0.44\textwidth}
  \caption{$Q_{ij}^v$ and $C_{ij}^v$ for buses.
   \label{fig:bus}}
\end{figure}

To compute $Q_{ij}^v$ and $C_{ij}^v$, we consider two cases of the vehicle type $v$.
First, for unpredictable vehicles (\ie, $v=0$), $Q_{ij}^0$ and $C_{ij}^0$ are estimated as $q_{ij}^0$  and $p_{ij}^0$, respectively.
However, in the case of buses (\ie, $v>0$), estimating $Q_{ij}^v$ and $C_{ij}^v$ is more complicated
because at intersection $i$, the outgoing edges created by type-$v$ bus become either
all available or all unavailable.
% known future trajectories of the buses make it possible to forward packets to the best neighbor intersection $j$
%among all augmented edges from the buses without any uncertainty.
Fig.~\ref{fig:bus} shows an example of such scenarios.
If a type-$v$ bus arrives at $i$, then all the corresponding outgoing edges become available,
in which case only the edge with the highest priority (under a decision $\bm u_i$) is used.
For instance, if the edge from $i$ to $l$ has the highest priority among all the edges of type-$v$ bus,
the bus carries packets to $l$, and hence,
$Q_{ij}^v$ and $C_{ij}^v$ for other edges (\ie, edges to $j,k,m$) become zero.
This shows that for $v>0$, $Q_{ij}^v$ and $C_{ij}^v$ depend on the routing decision $\bm u_i$.
Thus, if $e_{ij}^v$ is the best edge among all augmented edges from type-$v$ vehicles,
$Q_{ij}^v$ and $C_{ij}^v$ for $v>0$ are equal to $q_{i}^v$ and $p_{i}^v$, respectively.
Otherwise, $Q_{ij}^v$ and $C_{ij}^v$ are zero.
To describe this as an equation, we introduce a new notation $>_{\bm u_i}$
such that $e_{ij}^v >_{\bm u_i}e_{ik}^w$ if $e_{ij}^v$ has higher priority than $e_{ik}^w $ under a routing decision $\bm u_i$.
%To describe this as an equation, we define $R(\bm u_i,e_{ij}^v)$ as a function that returns the routing priority of an edge $e_{ij}^v$
%under the routing decision $\bm u_i$.
%Hence, a high value of $R(\bm u_i,e_{ij}^v)$ indicates high priority of $e_{ij}^v$.
The following summarizes the computation of $Q_{ij}^v$ and $C_{ij}^v$ for all types of vehicles:
\begin{eqnarray}
Q_{ij}^{v}(\bm u_i) = 
	\left\{
	\begin{array}{rl}
	
	q_{ij}^{v}	 & \text{if } v = 0 \\
	q_{i}^{v} 	& \mbox{if $v > 0$ and $e_{ij}^v >_{\bm u_i}e_{ik}^w$,}\\
			& \mbox{$\forall e_{ik}^{w} \in \mathcal{O}(i)$ s.t. $w=v$} \\
	0  			& \text{otherwise}\\
	\end{array}
	\right.\nonumber
\\
	C_{ij}^{v}(\bm u_i) = 
	\left\{
	\begin{array}{rl}
	
	p_{ij}^{v}	 & \text{if } v = 0 \\
	p_{i}^{v} 	 & \mbox{if $v > 0$ and $e_{ij}^v >_{\bm u_i}e_{ik}^w$,}\\
			& \mbox{$\forall e_{ik}^{w} \in \mathcal{O}(i)$ s.t. $w=v$} \\
	0  		& \text{otherwise}\\
	\end{array}
	\right. 
\label{eq:QC}	
\end{eqnarray}

%% file: 4_3_LinkCost.tex
%===================================== Link Delay ======================================
\subsection{Expected Delay on Edges} \label{subsection4_link_delay}
Recall that $d_{ij}^v$ is the expected data delay on an edge $e_{ij}^v \in \mathcal{L}$.
The delay $d_{ij}^v$ can be estimated using the average vehicle density and speed on $e_{ij}^v$ and the length of $e_{ij}^v$,
which can be easily obtained from the vehicle traffic statistics.
Note that if $e_{ij}^v$ corresponds to a set of multiple road segments in $\mathcal{R}$ (\ie, $e_{ij}^v \in \mathcal{R}'\setminus\mathcal{R}'_s$),
then on $e_{ij}^v$ the corresponding bus ``carries" the packets all the way to intersection $j$.
On the other hand, if $e_{ij}^v$ corresponds to a single road segment in $\mathcal{R}$ (\ie, $e_{ij}^v \in \mathcal{R}'_s$),
then V2V packet forwarding is allowed on $e_{ij}$ as discussed in Section~\ref{subsection4_overview}.
Hence, the delay on the edge is estimated differently depending on the number of hops in the edge.

\subsubsection{$d_{ij}^v$ on $e_{ij}^v \in \mathcal{R}'_s$}
Again, on this type of edges, a data packet is forwarded to another vehicle ahead.
Clearly, this V2V forwarding can significantly reduce the delay.
The delay $d_{ij}^v$ depends on the vehicle density $\rho_{ij}$ on $e_{ij}^v$
since if the density is high, then there is a high chance of V2V forwarding.
Note that if the WiFi transmission range $R$ is long, then there is also a high chance of V2V forwarding.
However, if a vehicle on $e_{ij}^v$ does not meet another vehicle in the transmission range,
then it has to carry its packets all the way to intersection $j$.
These factors can be integrated in several ways.
In this paper, we adopt the delay model in\cite{VADD} as follows:
\begin{align}
d^v_{ij} = (1-e^{-R \cdot \rho_{ij}}) \cdot \frac{l_{ij} \cdot c}{R} + e^{-R \cdot \rho_{ij}} \cdot \frac{l_{ij}}{s_{ij}}
\text{, for } v = 0\label{eq:link_cost}
\end{align} 
where $l_{ij}$, $s_{ij}$ and $c$ are the length of road segment $e_{ij}$, the average vehicle speed on $e_{ij}$ and the wireless transmission delay, respectively.
The first term in~(\ref{eq:link_cost}) is the expected delay contributed by V2V forwarding,
and the second term is the expected carrying delay.
The expression in~(\ref{eq:link_cost}) shows that the delay decreased as the transmission range $R$,
vehicle density or vehicle speed $s_{ij}$ increases.
%Note that $l_{ij}$ and $c$ represent the length of road segment $e_{ij}$ and the wireless transmission delay.

\subsubsection{$d_{ij}^v$ on $e_{ij}^v$  in $\mathcal{R}'\setminus\mathcal{R}'_s$}
In this case, packets are carried by the bus all the way.
Let $\mathcal{B}(e_{ij}^v)$ be the set of road segments between intersection $i$ and $j$ along the route of type-$v$ bus.
The packet delay on $e_{ij}^v$ depends only on the average speed of the bus with type $v$, denoted by $s_{ij}^v$, and the length of road segments in $\mathcal{B}(e_{ij}^v)$.
Hence, the expected delay on edge $e_{ij}^v$ can be estimated by the following equation:
\begin{align}
 d_{ij}^v =\sum_{e_{ij} \in \mathcal{B}(e_{ij}^v)} \frac{l_{ij}}{s^v_{ij}}\text{, for } v >0 \label{eq:link_cost_bus}
\end{align} 

\textbf{Remark:} The delay function clearly shows that our routing algorithm would prefer to forward packets along
the edges with high density and high average vehicle speed.

%Eq.~\eqref{eq:link_cost_bus} means the total expected delay on the all paths of the bus
%whose type is $v$ from the intersection $i$ to $j$.

%
%\textbf{The exptected data delay on the link is classified into two types as we explained former subsection.
%The first case is that  two intersections are neighbors.
%The second case is that two intersections are connected by bus.
%}
%\vspace{1cm}
%
%\textbf{ [The first case : link cost can be defined by average vehicles' speed and density]
%}
%\textbf{ [ Figure + Equation + Explanation ]
%}
%\vspace{8cm}
%
%\textbf{ [The second case : link cost can be estimated total length of the path (by bus) and average speed of
%bus ]
%equation will be added 
%[ Equation + Explanation ]}
%\vspace{5cm}

%% file: 5_Simulation.tex
\section{Performance Evaluation} \label{section_5}

In this section, 
we evaluate the performance of our routing algorithms based on real traces: Shanghai taxi\cite{ShanghaiTaxi} and 
bus\cite{BusTrace} trace.
The results show that our routing algorithms improve the delay performance by up to 105\% against the existing algorithm\cite{GPSR} in terms of the ratio of packets delivered to destinations in reasonable time.

%\subsection{Simulated Algorithm}
%%
%Our optimal algorithms can be distinguished as optimal algorithm-U or optimal algorithm-P.
%An optimal algorithm-U is a routing solution of MDP formulation which considers all vehicles as unpredictable vehicles
%when we cannot obtain future trajectories of any vehicles.
%An optimal algorithm-P is a MDP solution over an augmentedd road network graph $G'$ designed by knowing future trajectories of predictable vehicles (\eg, buses).
%%
%We compare our optimal algorithms to the Epidemic routing protocol\cite{DTN1_Vahdat} and GPSR\cite{GPSR}, an existing unicast routing algorithm.
%GPSR is a geographical routing algorithm in which a node forwards its packets to a node whose position is closer to a destination.
%%
%To apply GPSR in anycast routing, we assume that a forwarding decision between nodes is determined
%by comparing their geographic distance from the closest destination.
%%
%Note that other routing algorithms in unicast cannot be directly applied into anycast.
%%
%Epidemic routing is a multi-copy based routing protocol in which a node copies its packets to every node it encounters
%that does not have a copy.
%However, we evaulate this to show that our single-copy based routing protocol 
%overperforms since generating significantly less wireless congestions.

\subsection{Simulation Setup}

\begin{figure}[]
  \centering
  \epsfig{file=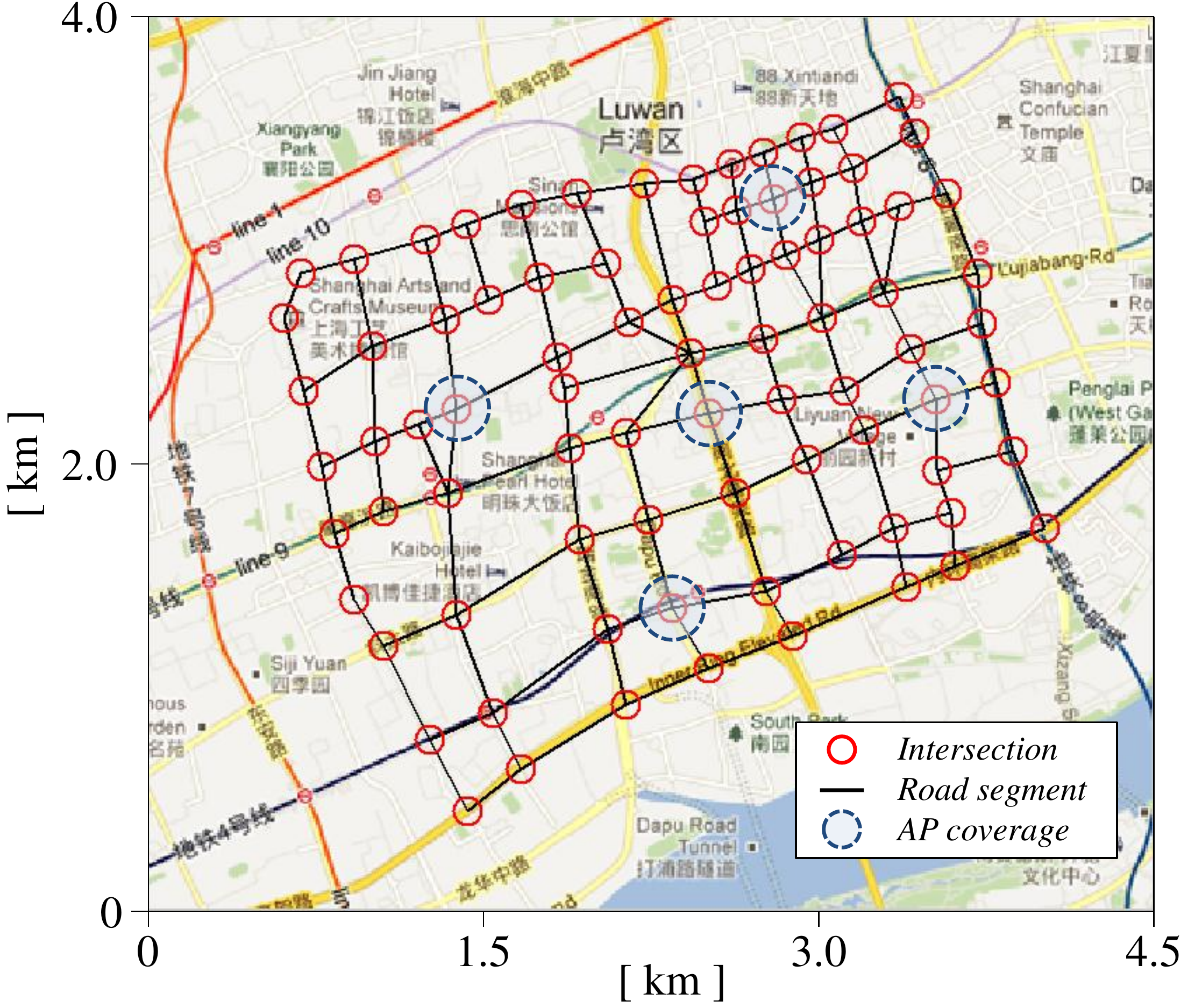,width=0.44\textwidth}
  \caption{Road network topology and AP deployment in Shanghai urban area
   \label{fig:map}}
\end{figure}

To verify our optimal routing algorithms, we use GPS traces of 
4800 taxis\cite{ShanghaiTaxi} and 2300 buses\cite{BusTrace} in Shanghai,
where the location information of each vehicle is recorded at every 30 seconds 
in 30km$\times$30km Shanghai for 28 days.
To focus on the sensing scenario of downtown area,
we select 4.5km$\times$4km Shanghai downtown, 
which consists of 84 intersections and 112 road segments, as shown in Fig.~\ref{fig:map}.
The selected area is modeled as a road network graph as discussed in Section~\ref{section_3}.
%To correct GPS errors in Shanghai trace, we use Google map\cite{GoogleMaps} that provides the exact locations 
%of intersections and roads.
Fig.~\ref{fig:map} also shows 5 candidate intersections where APs can be placed (see the intersections with dotted circle).
%We uniformly place APs at 5 intersections in the area as indicated in Fig.~\ref{fig:map}.
The algorithms will be evaluated for various numbers of APs.
We choose 120 taxis and 80 buses (6 types of buses) 
which have relatively low GPS errors and move within the area of our interest for more than 20 minutes during one hour period from 10AM to 11AM.
Since vehicles in Shanghai traces usually move outside the selected area,
we define ``effective number of vehicles" as the average number of vehicles which stay in the selected area during simulation,
and use it instead of the number of all vehicles when describing our simulation setup.
We denote by $N'$ the effective number of vehicles.
Note that $N'$ can be viewed as the vehicle density.

We implement the vehicular sensor network on a well-known wireless network simulator, GloMoSim\cite{glomosim}, 
using 802.11a MAC layer protocol\footnote{Although the last update year of GloMoSim was 2001, the IEEE 802.11a MAC protocol stack of GloMoSim
is still consistent to the current standards and many of physical layer models are included\cite{802.11ec}.}.
In the VSN, every vehicle moves along the road and senses the urban area.
The vehicles are assumed to generate sensing data packets when they satisfy at least one of the following conditions:
1) a vehicle moves 100m without any data generation.
2) a vehicle moves without any data generation during 30 seconds.
3) a vehicle reaches at one of 84 intersections.
Thus, vehicles generate packets based on their moving distance, time and geographic position.
All the vehicles and APs periodically send beacon packets to detect each other every second.
If they detect each other, they try to send their packets based on their routing algorithms.
Table~\ref{tbl:setting} presents parameters and scenarios.

%\begin{figure*}[ht h!]
%  \centering
%% \subfigure[GPSR]{\epsfig{file=figures/graph6.pdf,width=0.44\textwidth}\label{fig:coverage1}}
%%  \subfigure[OVDF-U]{\epsfig{file=figures/graph6.pdf,width=0.44\textwidth}\label{fig:coverage2}}
%% \subfigure[GPSR]{\epsfig{file=figures/graph6.pdf,width=0.44\textwidth}\label{fig:coverage1}}
%%  \subfigure[OVDF-U]{\epsfig{file=figures/graph6.pdf,width=0.44\textwidth}\label{fig:coverage2}}
%  \subfigure[GPSR]{\epsfig{file=figures/graph1_1.pdf,width=0.40\textwidth}\label{fig:coverage1}} \hspace{1cm}
%  \subfigure[OVDF-U]{\epsfig{file=figures/graph1_2.pdf,width=0.40\textwidth}\label{fig:coverage2}}
%  \subfigure[OVDF-P]{\epsfig{file=figures/graph1_3.pdf,width=0.40\textwidth}\label{fig:coverage3}} \hspace{1cm}
%  \subfigure[Epidemic]{\epsfig{file=figures/graph1_4.pdf,width=0.40\textwidth}\label{fig:coverage4}}
%  \caption{Vehicular sensing coverage in the scenario of 3 APs and 80 effective vehicles (including taxies and 6 different types of buses).
%  \label{fig:coverage}}
%\end{figure*}

\begin{figure*}[ht h!]
  \centering
% \subfigure[GPSR]{\epsfig{file=figures/graph6.pdf,width=0.44\textwidth}\label{fig:coverage1}}
%  \subfigure[OVDF-U]{\epsfig{file=figures/graph6.pdf,width=0.44\textwidth}\label{fig:coverage2}}
% \subfigure[GPSR]{\epsfig{file=figures/graph6.pdf,width=0.44\textwidth}\label{fig:coverage1}}
%  \subfigure[OVDF-U]{\epsfig{file=figures/graph6.pdf,width=0.44\textwidth}\label{fig:coverage2}}
  \subfigure[GPSR]{\epsfig{file=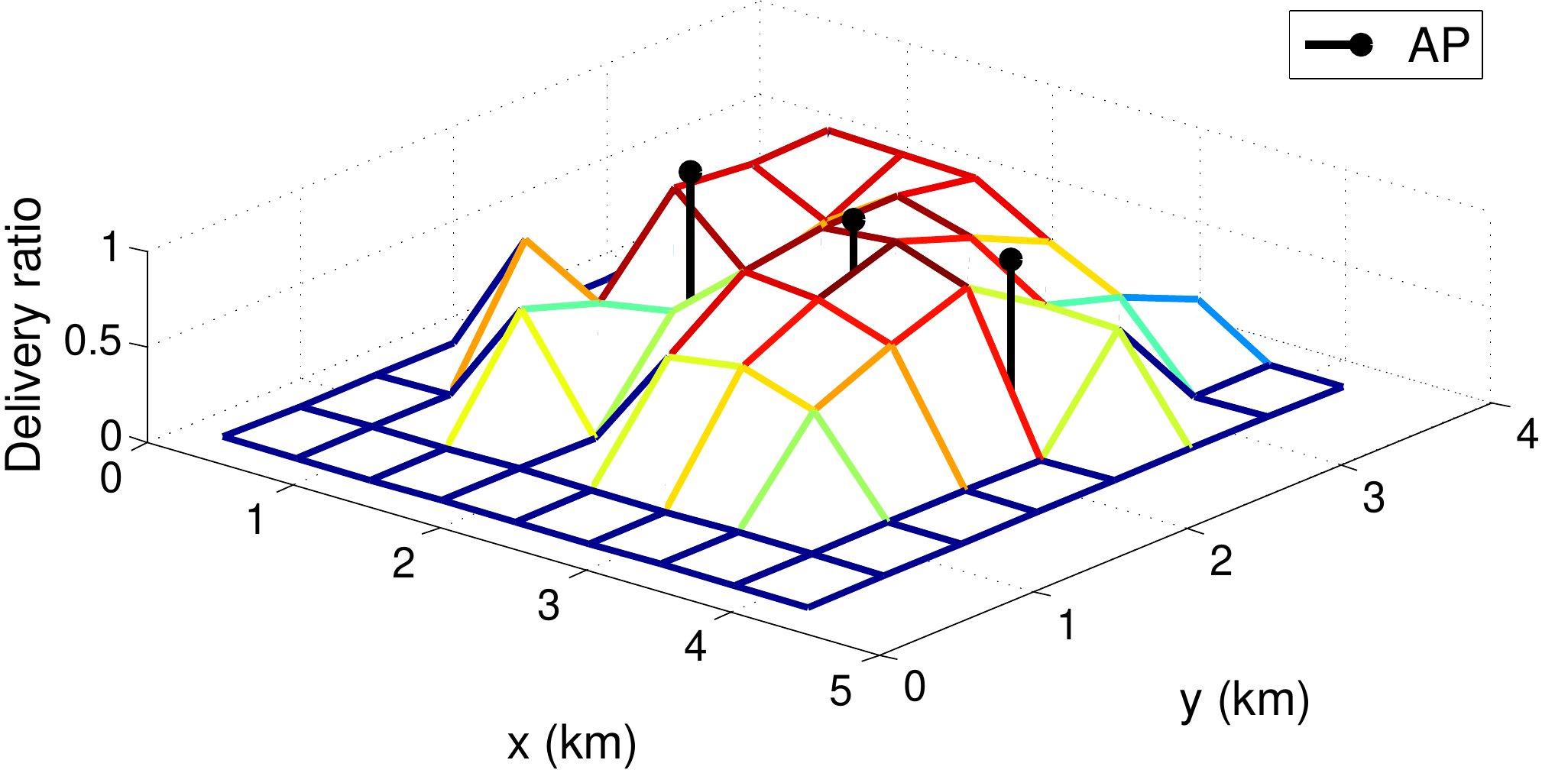,width=0.42\textwidth}\label{fig:coverage1}} \hspace{1cm}
  \subfigure[OVDF-U]{\epsfig{file=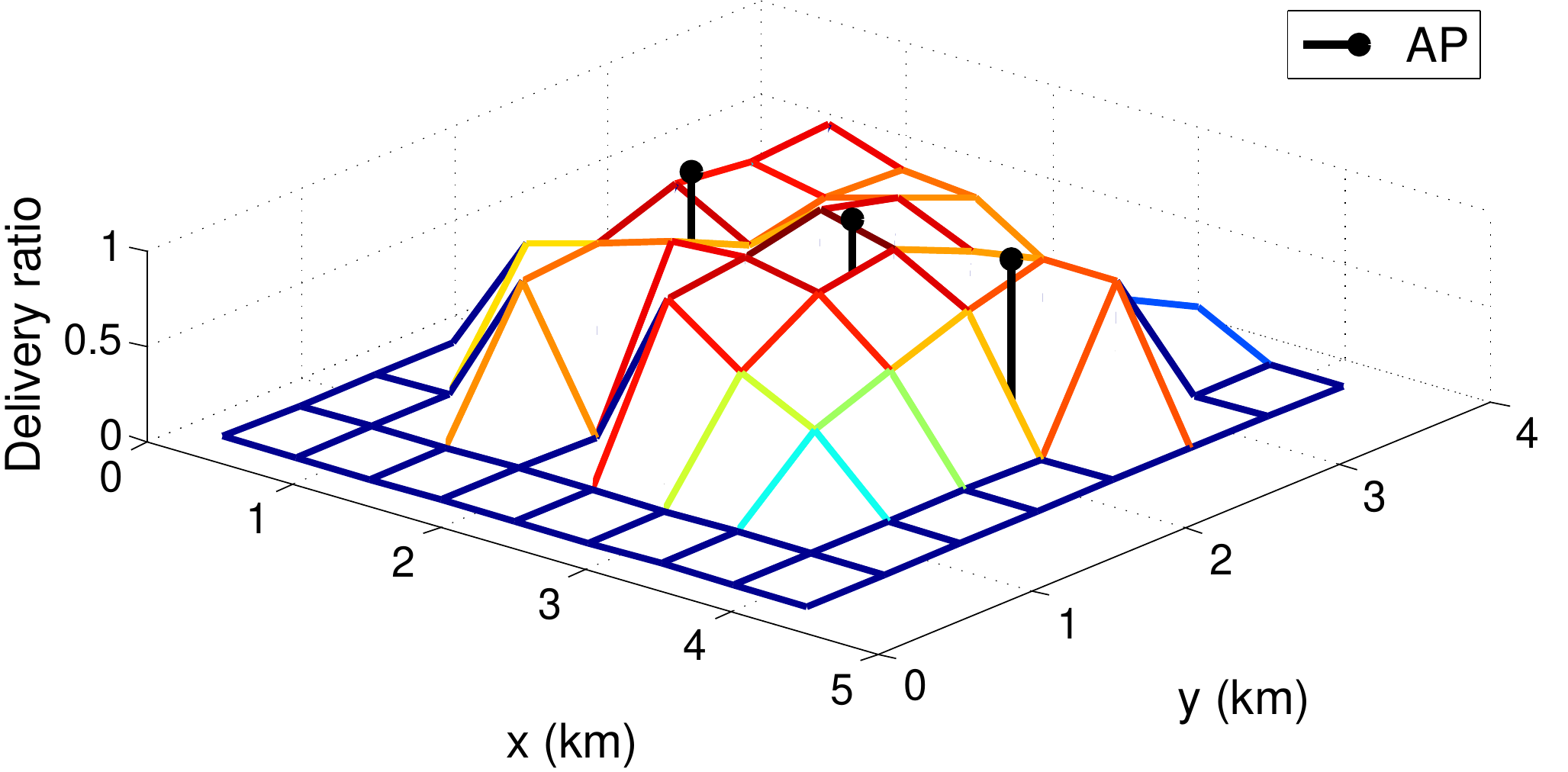,width=0.42\textwidth}\label{fig:coverage2}}
  \subfigure[OVDF-P]{\epsfig{file=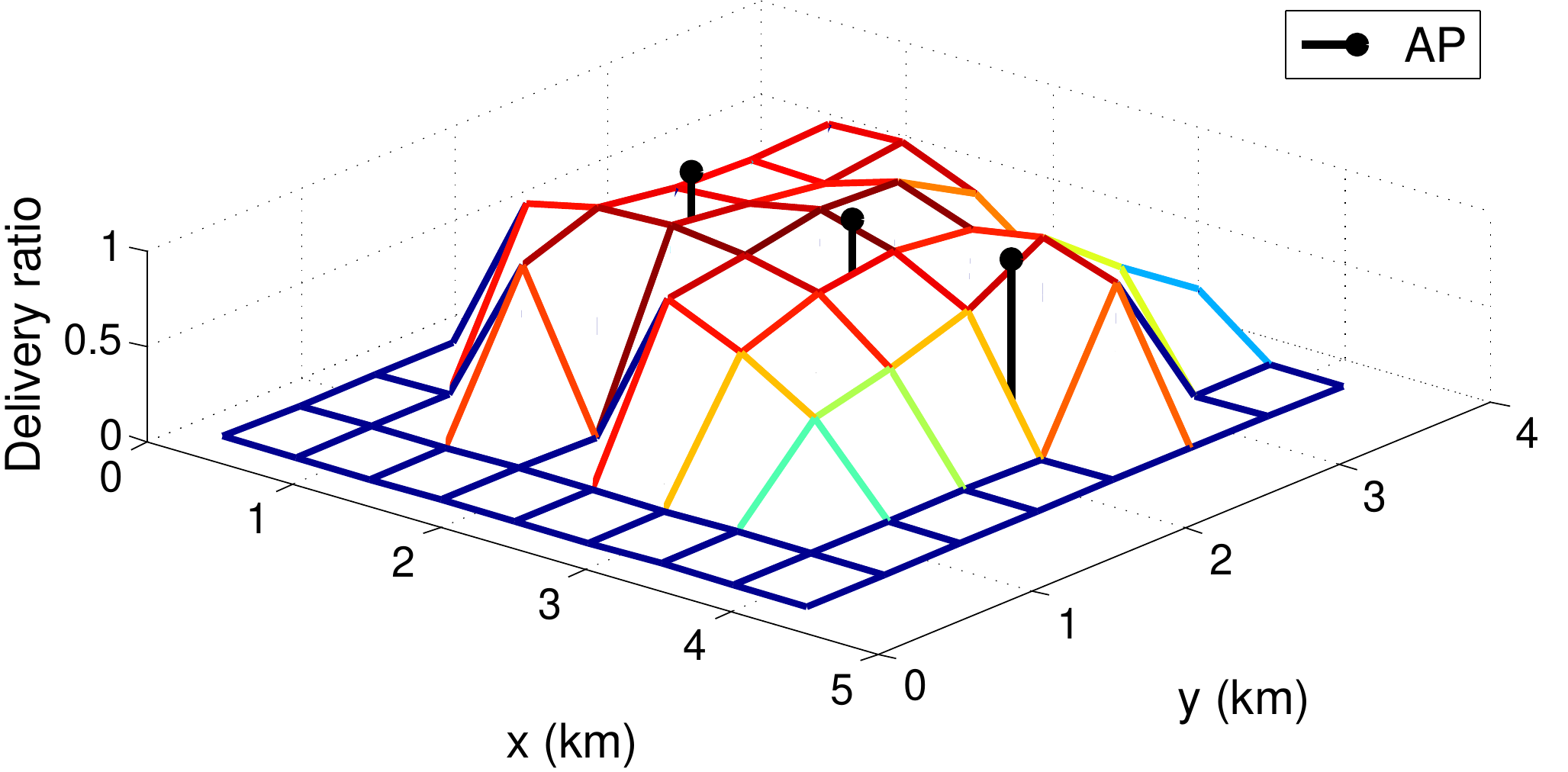,width=0.42\textwidth}\label{fig:coverage3}} \hspace{1cm}
  \subfigure[Epidemic]{\epsfig{file=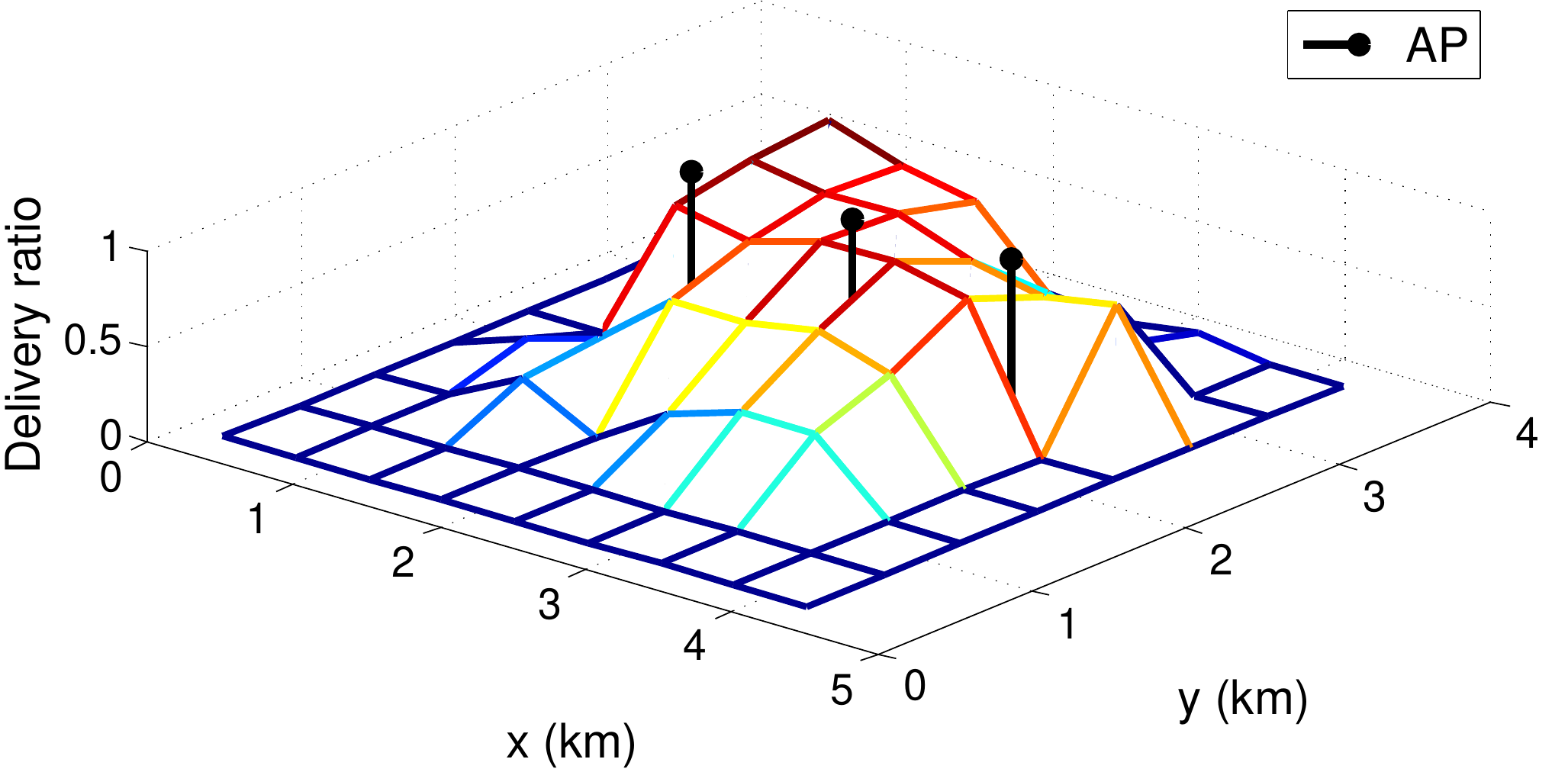,width=0.42\textwidth}\label{fig:coverage4}}
  \caption{Vehicular sensing coverage of four tested algorithms : (a) GPSR, (b) OVDF-U, (c) OVDF-P, (d) Epidemic.  
  In all cases, the number of AP is 3 and the number of effective vehicles is 80 (including taxis and 6 different types of buses).
  \label{fig:coverage}}
\end{figure*}

\begin{table}[ t! h!]
  \caption{Simulation Settings}
  \begin{centering}
    \small
    \tabcolsep = 0.25in 
    \begin{tabular}{|c|c|}
      \hline
      Parameter Name & Definition\\%tabularnewline
      \hline
      \hline
      Effective number of & 95(30), 80(27), 65(22),\\
      vehicles (buses) & 55(18), 45(14)\\
      \hline
	Number of bus lines & 6\\
      \hline
	Number of APs & 1, 3, 5\\
	\hline
	Simulation time & 1 hour\\
\hline
	Wireless device & 802.11a\\
\hline
	Data packet size & 512 Bytes\\
\hline
	Communication range & 150m\\
      \hline
    \end{tabular}\label{tbl:setting}
    \par\end{centering}
\end{table}

\subsection{Tested Algorithms}
We test two VSN routing algorithms in the literature: 1) Epidemic routing protocol\cite{DTN1_Vahdat} which floods packets over the network and 2) GPSR\cite{GPSR}
which forwards packets to a node located closer to a destination.  
In anycast version of GPSR, the forwarding decision at an intersection is made by comparing each neighbor intersection's distance
from its cloest AP.
%between nodes is determined by comparing their geographic distance from the closest destination.
Further, our version of GPSR adopts buffer in each vehicle, so that when vehicles fail to transmit packets, they store the packets in the buffer and keep moving around.
We consider two versions of our optimal routing algorithm:
(i) Optimal VSN Data Forwarding that regards all the vehicles as those with unpredictable trajectories (denoted by OVDF-U),
and (ii) Optimal VSN Data Forwarding that takes into account, the vehicles with known future trajectories (denoted by OVDF-P).
The forwarding decision in OVDF-U is computed assuming that the vehicle type set $\mathcal{V} = \{0\}$, even if
there are buses moving in the network.
On the other hand, the algorithm OVDF-P fully exploits the known trajectories by incorporating them into the formulation
through the augmented network graph.
We will study the effect of utilizing the vehicles with known trajectories.
Note that other routing algorithms for unicast cannot be directly applied to the anycast setting.

\input{5_1_Coverage}
\input{5_2_Density}

\input{5_3_Packets}

%% file: 5_1_Coverage.tex
\subsection{Simulation Results}

\subsubsection{Sensing Coverage}
\begin{figure*}[ht h!]
  \centering
  \subfigure[1 AP]{\epsfig{file=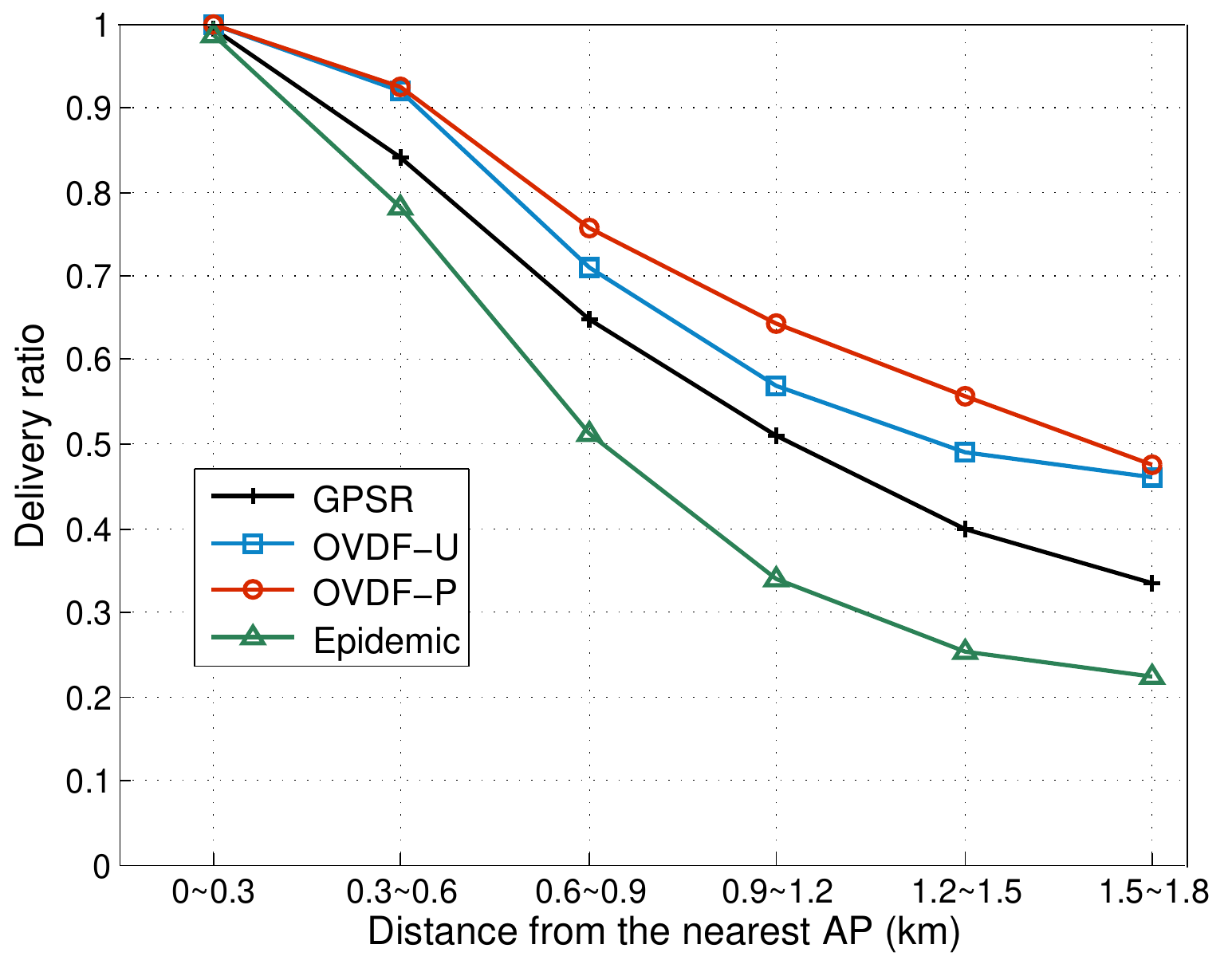,width=0.323\textwidth}\label{fig:distance1}}
  \subfigure[3 AP]{\epsfig{file=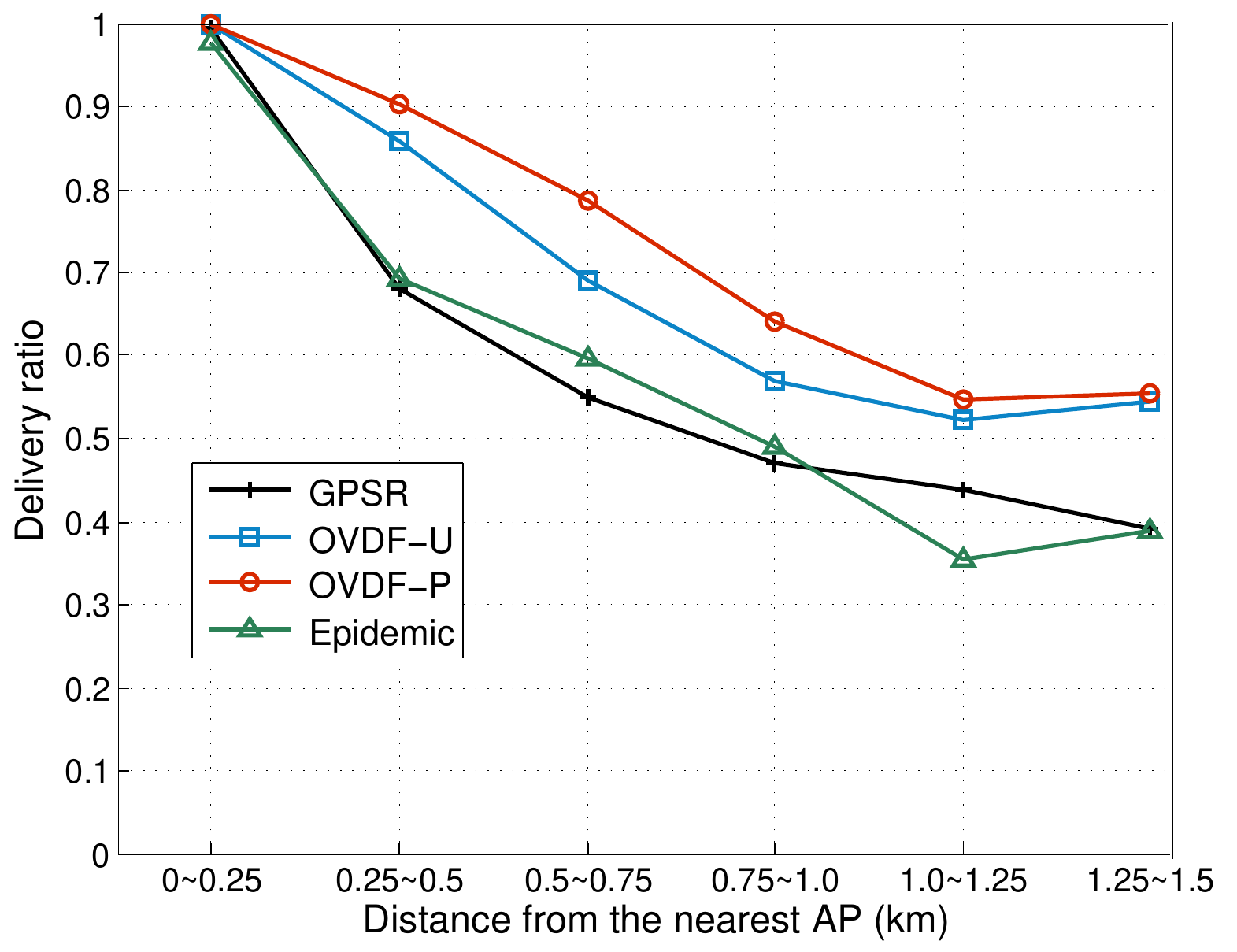,width=0.328\textwidth}\label{fig:distance2}}
  \subfigure[5 AP]{\epsfig{file=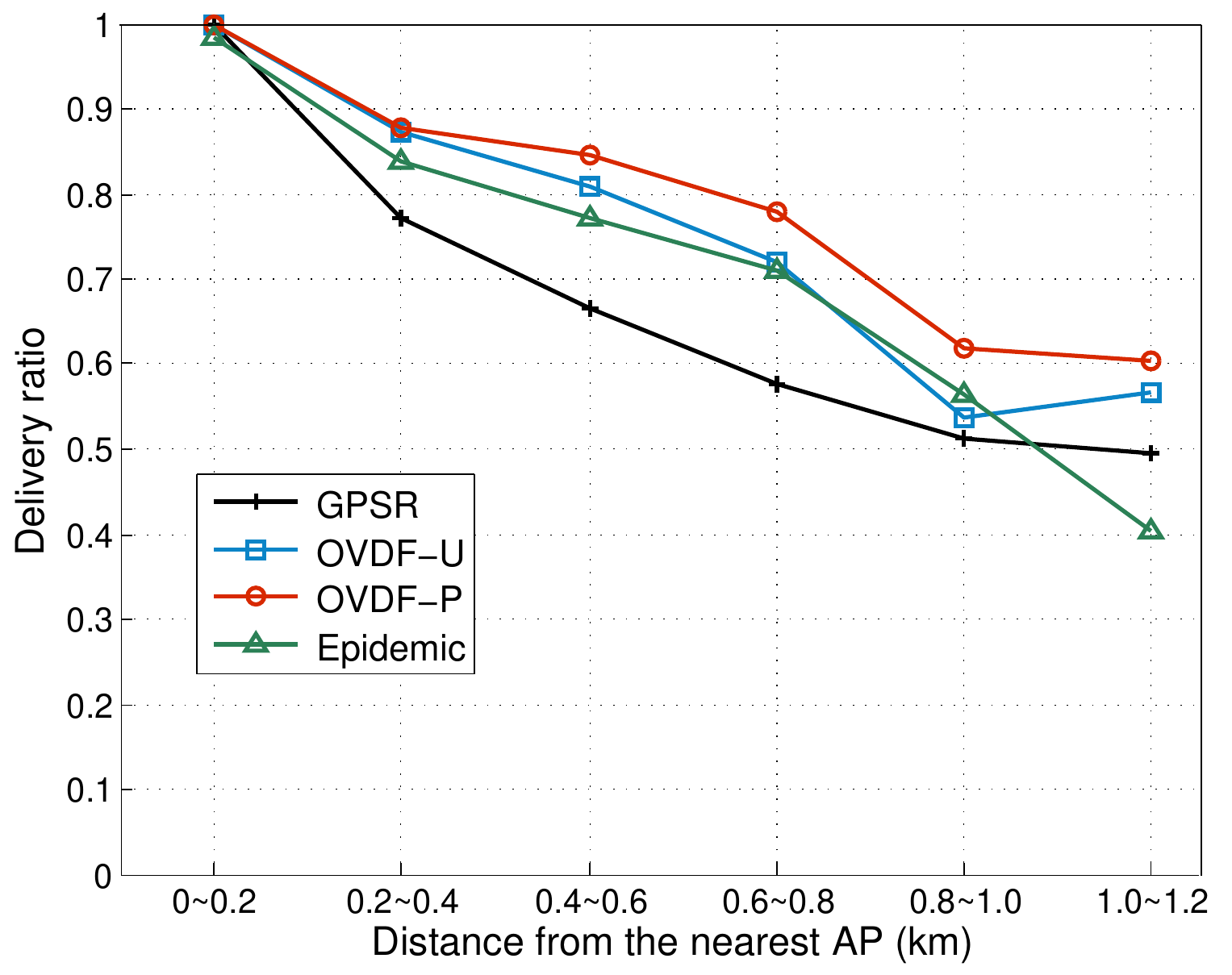,width=0.323\textwidth}\label{fig:distance3}}
  \caption{Delivery ratio with 10 minutes deadline versus distance to AP for the various number of APs: (a) 1 AP, (b) 3 AP, (c) 5 AP. In all cases, the number of effective vehicles is 55 (including taxis and 6 different types of buses).
  \label{fig:distance}}
\end{figure*}

%\begin{figure*}[ht h!]
%  \centering
%  \subfigure[1 AP - 3/3]{\epsfig{file=figures/graph3_1.pdf,width=0.32\textwidth}\label{fig:density1}}
%  \subfigure[3 AP - 3/3]{\epsfig{file=figures/graph3_2.pdf,width=0.32\textwidth}\label{fig:density2}}
%  \subfigure[5 AP - 3/3]{\epsfig{file=figures/graph3_3.pdf,width=0.32\textwidth}\label{fig:density3}}
%  \caption{Data delivery ratio within 10 minutes versus distance to AP from their generated position in a scenario where xxx vehicles (including yyy buses) moves through Shanghai urban street with generating kkkkk sensing data.
%  \label{fig:density}}
%\end{figure*}

We first evaluate the sensing coverage of the four algorithms mentioned above.
%In the VSN applications such as Intelligent Transportation System (ITS), 
%the information that passed a certain deadline is usually ignored, and
%it is desirable for a new packet to be delivered no later than the deadline.
One of the most important performance metrics in VSN routings is the delivery ratio within a certain deadline.
In our simulation, the deadline of a packet is fixed to 10 minutes since its creation.
%which was sensed within 10 minutes are generally allowable in the VSN services
%and other data with more than 10 minutes delay are ignored.
%If the delivery ratio is high, 
%VSN applications can provide more reliable service with
%the sufficient amount of sensing information.
%Thus, we quantify the sensing coverage by 
%presenting the 10-minute delivery ratio from each area.
To show the spatial coverage performance,
we divide 4.5km $\times$ 4km of Shanghai downtown
into a grid of 72 0.5km $\times$ 0.5km squares
and measure the delivery ratios of the algorithms for each square.
Those are 80 vehicles (including 27 buses) and 3 APs.

Fig.~\ref{fig:coverage} plots the delivery ratio within 10 minutes 
where the $x$-$y$ plane represents the grid and $z$-axis represents the delivery of sensing data packets
generated in the corresponding squares.
Out of all the total packets generated, more than 90\% of them are from 28 squares
(this is due to the road structure of Shanghai downtown imbalance among squares).
Note that the areas that generated too few packets do not give meaningful results.
Thus, we only consider those squares in the results and call them the ``valid squares.''

Compared to GPSR and Epidemic routing,
OVDF-U and OVDF-P show higher packet delivery ratios in most of the regions.
Especially, as shown in Fig.~\ref{fig:coverage1} and Fig.~\ref{fig:coverage3},
11 squares (out of 28 valid squares) in OVDF-P achieve at least 20\% higher delivery ratio (up to 105\% higher) than those in GPSR, and 
on average, the delivery ratio gain over GPSR is 22\% for all of the valid squares.
%the delivery ratios at all of valid grids are greater than those in GPSR.
Compared to OVDF-U, OVDF-P shows 9\% higher average delivery ratio,
and thus, OVDF-P guarantees the best sensing coverage in this scenario.
In Fig.~\ref{fig:coverage4}, Epidemic routing shows the lowest packet delivery ratios in most of the squares,
due to heavy congestion caused by packet replications.

%
%Compared to Epidemic routing in Fig.~\ref{fig:coverage4}, 
%OVDF-P shows higher delivery ratios at 19 of 28 grids, and the maximum gain of delivery ratio is up to 30\%.

% --> more performance explanation 

% Distance versus Performance
An important observation in Fig.~\ref{fig:coverage} is that, for any routing algorithm, most of the packets generated in the regions close to APs 
are delivered to an AP within 10 minutes.
This is because those data are close enough to be forwarded to the AP in one or a small number of hops, and thus,
the impact of optimal routing decisions on the delivery ratio is relatively small.
On the other hand, the intelligence of routing algorithms have a huge impact on 
the delivery ratio for data packets created in the edge areas that are far from APs.
Note that there are abundant routes from an edge area to the APs,
however only a few of them are those with low delay. 
Consequently, in the edge areas, the delay performance can vary substantially depending on the routing algorithms.
Hence to verify the performance of the routing algorithms,
it is necessary to evaluate the delivery ratio from distant areas.
In following, we study the routing performance against the distance to APs.

%\textbf{because on the way to an AP, those packets are more likely to be forwarded to the regions where there are few vehicles,
%in which case the delay can significantly increase.}
%
%Hence, it is important to study the routing performance against distance to APs.
%Note that the delivery ratio of edge areas are relatively low
%and the timely delivery from those areas is also important to achieve high sensing coverage.
%Since the delivery ratios are shown to depend on the distance between the areas and APs,
%we further classify grids to the distances between them and the nearest AP.

%Thus, a good routing algorithm for VSNs should decrease delivery delays 
%of those packets so that a large portion of them is delivered to one of 
%APs within allowable time.

%However, delivery ratios of data packets created at edge areas which are far from APs are relatively low,
%and greatly changed depending on routing algorithms.

%

\subsubsection{Performance against Distance to APs}
\label{sec:perf_dist}
%We plot the delivery ratio of sensing packets against 
%the distance between the grid where they are generated and the nearest AP.
%we plot the delivery ratio against the distance between the sensing grid and the nearest AP.
%we evaluate the delivery performance on the packets by a distance to the nearest AP from a position where they are generated.

The packet delivery ratio against the distance is estimated as follows:
Consider a packet generated at a position $X$ in the plane,
and let $Y$ be the position of the nearest AP from $X$.
If the distance between $X$ and $Y$ is the range of $a\sim b$ km,
then the delivery of the packet contributes to the delivery ratio
from the areas whose nearest APs are $a\sim b$ km away.
There are 55 effective vehicles including 18 buses.

Fig~\ref{fig:distance} shows that for the regions within the wireless communication range of APs, the delivery ratio is 100\%.
As the distance between grids and APs increase, the delivery ratio decreases.
Note that OVDF-P outperforms GPSR by up to 40\%, 43\% and 37\% 
for the scenario of 1 AP, 3AP and 5 APs, respectively.
For the entire range of the distance, the performance gap between GPSR and OVDF-U is larger in anycast scenarios (\ie, 3 or 5 APs) than in the unicast scenario (\ie, 1 AP).
This is because GPSR is not designed for anycast routing and thus cannot fully exploit the advantage of multiple destinations.
Moreover, OVDF-U shows higher delivery ratio than GPSR even in the unicast scenario
since we estimate the expected delay on a road segment based on not only its length but also the vehicle density
and speed on the road.
Clearly, OVDF-P shows the best delivery ratio performance among all the algorithms, especially for the packets generated far from APs.

As shown in Fig.~\ref{fig:distance1}, in the unicast scenario,
Epidemic routing shows poor delivery performance 60\% lower than that of OVDF-P at the distance of 1.2km.
In anycast scenarios, spreading the packets all over the network can significantly improve the delay performance
since there are multiple APs throughout the network.
On the other hand, such an effect becomes marginal in the unicast scenario where there is only one AP.
Moreover, it incurs severe packet collisions as many heavily-loaded vehicles would attempt to transmit to the single AP.
This is why Epidemic routing shows poor performance in the unicast scenario.

%
%It means that GPSR cannot fully exploit advantages came from the existence of multiple destinations since GPSR is not designed for anycast routing.
%Moreover, OVDF-P and OVDF-U shows higher delivery ratios than GPSR even in an unicast scenario
%since we estimate the expected delay on road segment based on not only length of the road but also 
%vehicles density and speed on the road.
%Clearly, OVDF-P shows the best delivery performance among all algorithms, especially in the case where packets are far from APs.
%As shown in Fig.~\ref{fig:distance1}, in a unicast scenario, Epidemic routing shows poor delivery performance \ie, 60\% lower than that of OVDF-P at the distance of 2km.
%This is because, in anycast scenarios data traffics generated by Epidemic routing are spread to each destination
%but, in case of a unicast scenario, the flooding does not work efficiently but incurs a large amount of data collisions near an AP.
%Due to the lower delivery ratio of longer distance areas,
%we only consider the delivery ratio of sensing packets generated at those area in the following results.
%

%% file: 5_2_Density.tex
\subsubsection{Performance against the Number of Vehicles}
%\begin{figure*}[ht]
%  \centering
%  \subfigure[1 AP - 3/3]{\epsfig{file=figures/graph3_1.pdf,width=0.28\textwidth}\label{fig:distance1}}
%  \subfigure[3 AP - 3/3]{\epsfig{file=figures/graph3_2.pdf,width=0.28\textwidth}\label{fig:distance2}}
%  \subfigure[5 AP - 3/3]{\epsfig{file=figures/graph3_3.pdf,width=0.28\textwidth}\label{fig:distance3}}
%  \caption{Data delivery ratio within 10 minutes versus distance to AP from their generated position in a scenario where xxx vehicles (including yyy buses) moves through Shanghai urban street with generating kkkkk sensing data.
%  \label{fig:overhead}}
%\end{figure*}

\begin{figure}[]
\centering
  \epsfig{file=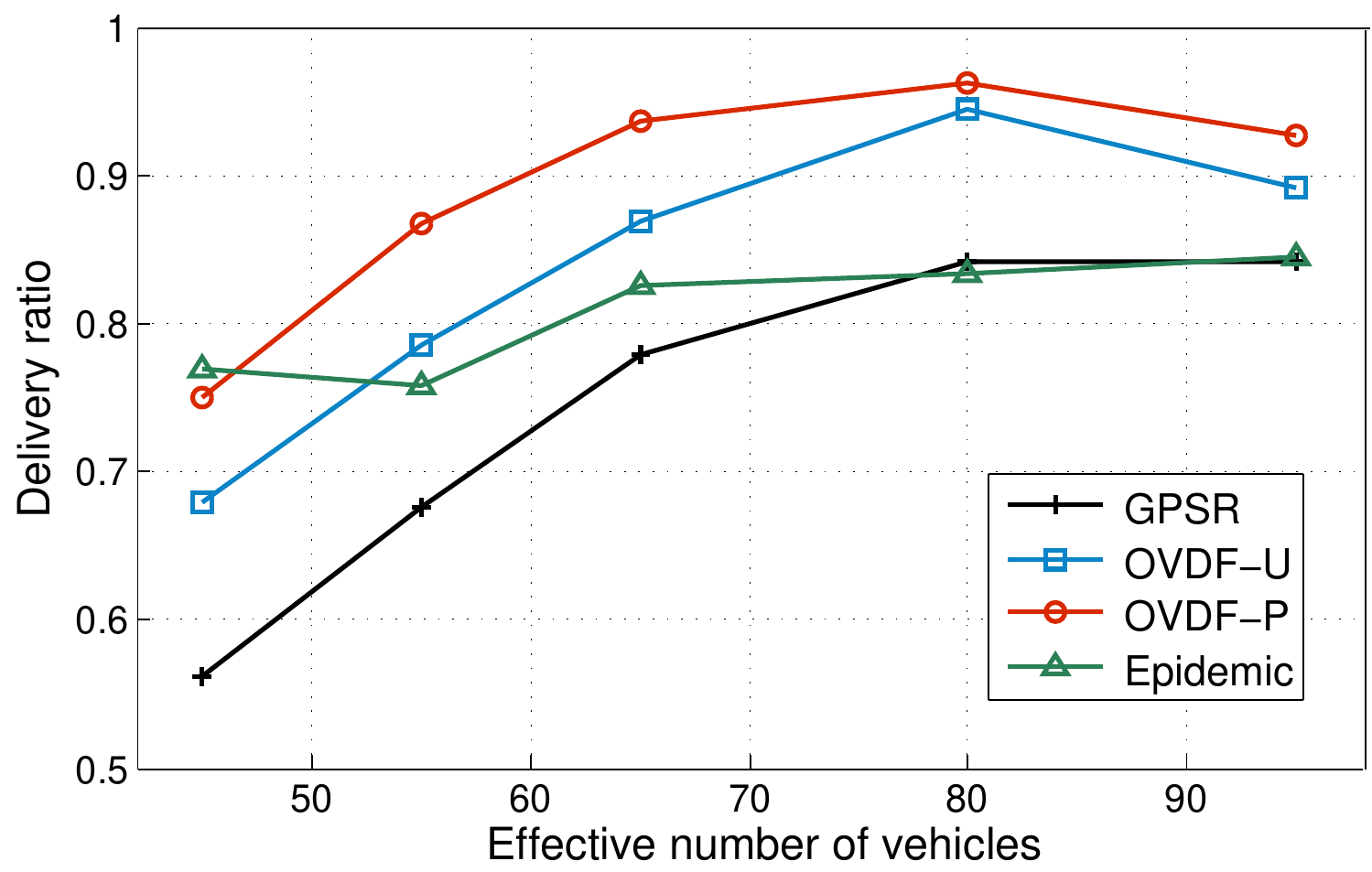,width=0.43\textwidth}\label{fig:density1}
  \caption{Data delivery ratio within 10 minutes versus the effective number of vehicles when the number of AP is 5. 
  \label{fig:density}}\vspace{-0.2cm}
\end{figure}

Next, we compare the performance by changing the number of effective vehicles.
The number of APs is 5, and only the packets generated at least 400m away from all the APs are considered.
Fig.~\ref{fig:density} shows the delivery ratio of each algorithm.
The observation from these results are threefold.
First, when the density of vehicles in the road network is high,
the algorithms achieve comparable performance because,
in the densely connected vehicular network, packets are delivered to the destination mostly by V2V forwarding.
Second, the performance gap between OVDF-P and other algorithms increases as the density of vehicles decreases.
Fig.~\ref{fig:density} shows that OVDF-P achieves 11\% higher delivery ratio than OVDF-U, 
when the effective number of vehicles is $45$.
In sparse vehicular networks, the packet delivery depends heavily on carrying,
and thus the effect of vehicles with known trajectories is much more appreciated.
Third, in the case of low vehicle density, the delivery ratio of Epidemic routing is relatively high as opposed to the high density scenario
because when there are not many routes to destinations due to a small number of vehicles,
replicating the packets would greatly improve the chance of reaching a destination.

%% file: 6_Conclusion.tex
\section{Concluding Remarks}\label{section_6}

Many of emerging VSN applications require timely delivery of sensing data and wide sensing coverage.
However, this is a challenging problem in the VSN where the data links are intermittently connected.
To address the issues, 
we develop a delay-optimal VSN routing algorithm, capturing three key features in urban VSNs:
(i) vehicle traffic statistics, (ii) anycast routing and 
(iii) known future trajectories of vehicles such as bus. 
Using real traces of 120 taxis and 80 buses in Shanghai
we conduct extensive simulations on GloMoSim simulator,
and show that our optimal algorithm outperforms other existing algorithms.
Our results demonstrate that carefully designed packet routing algorithms
can greatly improve the delay performance in the VSN,
and thus are the key to the success of VSN applications
that require stringent delay performance guarantee.
In this paper, we focused single-copy routing algorithms.
Although multi-copy routings can incur serious congestion, there are scenarios
where multi-copy routings outperform single-copy routing.
Therefore, it would be interesting to extend our framework to account for multi-copy routing,
and we leave this as future study.